\documentclass[a4paper,twocolumn,10pt,unpublished]{quantumarticle}
\pdfoutput=1
\usepackage[numbers,sort&compress]{natbib}

\usepackage[english]{babel}
\usepackage{graphicx} 
\usepackage{physics}
\usepackage{url}
\usepackage{wasysym}
\usepackage[utf8]{inputenc}
\usepackage{algorithm}      
\usepackage{algpseudocode}
\usepackage{setspace}

\usepackage{xcolor}
\definecolor{darkred}{rgb}{0.6,0.05,0.05}
\definecolor{darkgreen}{rgb}{0.05,0.6,0.05}
\definecolor{darkblue}{rgb}{0.05,0.05,0.6}
\usepackage[colorlinks]{hyperref}
\hypersetup{colorlinks,
linkcolor=darkred,
filecolor=darkgreen,
urlcolor=darkblue,
citecolor=darkgreen}

\usepackage{amssymb}
\usepackage{mathbbol}
\usepackage{dsfont}

\usepackage{tikz}

\newcommand{\LL}{\mathcal{L}}

\algnewcommand\algorithmicgoto{\textbf{go to}}
\algnewcommand{\Goto}[1]{\State \algorithmicgoto~line~\ref{#1}}

\usepackage[normalem]{ulem}

\newcommand{\na}[1]{\overline{#1}}

\begin{document}


\title{Liouvillian spectral snapshots: A reduced-subspace approach for time-dependent and noisy 
open quantum systems}

\author{Hans Christiansen}
\affiliation{Niels Bohr Institute, University of Copenhagen, 2100 Copenhagen, Denmark}
\email{hans.christiansen@nbi.ku.dk}
\affiliation{Alice \& Bob, 53 boulevard du G\'{e}n\'{e}ral Martial Valin, 75015 Paris, France}
\author{André Melo}
\affiliation{Alice \& Bob, 53 boulevard du G\'{e}n\'{e}ral Martial Valin, 75015 Paris, France}
\author{Jens Paaske}
\affiliation{Niels Bohr Institute, University of Copenhagen, 2100 Copenhagen, Denmark}
\author{Fabrizio Minganti}
\affiliation{Alice \& Bob, 53 boulevard du G\'{e}n\'{e}ral Martial Valin, 75015 Paris, France}
\email{fabrizio.minganti@gmail.com}

\date{October 2026}

\begin{abstract}
Driven-dissipative quantum devices are routinely operated with time-dependent drives and exposed to temporally correlated classical noise. When these perturbations vary on timescales comparable to the system's intrinsic dynamics, rotating-wave and Markovian approximations can fail, making effective static descriptions unreliable. As a result, accurately modeling systems in this regime often relies on simulating the full time-dependent master equation. This is particularly demanding for processes requiring long integration times, such as rare switching between metastable states, or for stochastic perturbations requiring averaging over many noise realizations. To address these challenges, we introduce Liouvillian spectral snapshots (LSS), a reduced-basis method that enables efficient simulation of such systems by projecting the master equation onto a low-dimensional subspace capturing their coarse-grained dynamics. LSS constructs this subspace by computing the slowly decaying, low-frequency eigenoperators of the Liouvillian at a few adaptively sampled perturbation values and compressing them into a common basis. This basis is built once and can be reused across noise realizations and perturbation parameters, reducing the cost of ensemble averages and parameter sweeps. 
We benchmark LSS against exact dynamics on two examples. For a bistable Kerr resonator under two-tone driving and correlated frequency noise, a basis ten times smaller than the Liouville space reproduces the switching rates to within a few percent and reveals a strong, non-perturbative enhancement of switching when the perturbation is resonant with a long-lived oscillating mode. For an eleven-qubit repetition code under continuous error correction and collective $1/f$ phase noise, LSS accurately reproduces the dynamics using only $37$ basis vectors, compared with the full Liouville-space dimension of $4^{11}\approx4\times10^{6}$.
\end{abstract}

\maketitle

\section{Introduction}

Controlling quantum systems often relies on time-dependent drives and pumping mechanisms to steer their dynamics toward desired states~\cite{koch_2022,blais_circuit_2021,krantz_quantum_2019,bruzewicz_2019,burkard_2023}.
However, in experimental settings, parasitic interactions, dissipation, and temporally correlated noise perturb the intended evolution~\cite{sank_2016,paladino_2014,huelga_2013}.
Typical examples of these \textit{time-dependent perturbations} include $1/f$ and telegraph-like fluctuations that limit coherence in superconducting circuits~\cite{paladino_2014}, semiconductor spin qubits~\cite{yoneda_2018,burkard_2023}, and color centers~\cite{wolfowicz_2021}. In molecular and excitonic systems, structured vibrational environments also perturb the controlled dynamics~\cite{chin_2013,huelga_2013}. 
Flux noise in circuit quantum electrodynamics (cQED) provides a concrete example of these temporally correlated perturbations. Spectroscopic measurements reveal power spectral densities that scale approximately as \(1/f\) over 
a broad frequency range
~\cite{bylander_noise_2011,yan_flux_2016}, with additional Lorentzian contributions associated with strongly coupled two-level systems (TLS)~\cite{Yan2013}.
This structured flux noise limits the performance of a wide range of quantum devices and protocols, including the quantum efficiency of single-microwave photon sources~\cite{lu_quantum_2021}, coherence in flux-sensitive qubits~\cite{yoshihara_decoherence_2006,ithier_decoherence_2005}, and the lifetime of coherent states in parametrically driven systems~\cite{beaulieu_criticality-enhanced_2025,Reglade2024,Marquet2024}.

Predicting how quantum systems respond to time-dependent drives and noise requires models that capture the temporal correlations and spectral features of these perturbations.
This is particularly challenging for systems that support a manifold of slowly decaying processes alongside rapidly decaying modes~\cite{macieszczak2016,Minganti2018}.
The full equations of motion must then be integrated over long times while resolving the fast dynamics induced by the perturbation.
This situation arises, for example, when simulating \textit{rare events}, such as activated switching between metastable states~\cite{dykman1990}, the Kramers barrier-crossing problem in chemical physics~\cite{kramers1940,hanggi1990,bolhuis2002}, bit- and phase-flip errors in protected codes~\cite{lescanne2020,leghtas2015,putterman_2025,mirrahimi2014cat}, and, more generally, long-time relaxation~\cite{BucaProsen2012,manzano2014}.

Motivated by these experimental and theoretical challenges, we introduce the \textit{Liouvillian spectral snapshots} (LSS) method, illustrated in Fig.~\ref{fig:scheme}, to efficiently simulate the coarse-grained dynamics of systems subject to time-dependent perturbations.
LSS samples the Liouvillian at representative values of the perturbation and identifies the eigenoperators associated with slow evolution at each value.
These static spectral ``snapshots'' are then combined into a single reduced basis.
In practical applications, a few snapshots are often sufficient to span the slow subspace across the range of the perturbation.
Indeed, in the examples studied below, LSS finds a reduced subspace whose dimension is one to several orders of magnitude smaller than the full Liouville space.
The coarse-grained dynamics of the system can then be efficiently simulated by projecting the time-dependent master equation onto this reduced subspace.
The same reduced basis can be reused across many noise realizations, making LSS particularly appealing for ensemble averaging, enabling efficient computation of noise-averaged relaxation rates or observables.
Because the basis is built from the distribution of the perturbation rather than from its power spectrum, the same basis can be reused for broadband $1/f$ or white backgrounds, sharp Lorentzian peaks, and their combinations, allowing a direct comparison of noise of equal variance but different spectral content.

The paper is organized as follows. In Sec.~\ref{sec:method}, we present the LSS method in detail.
We then demonstrate its accuracy and efficiency through two applications.
The first, described in Sec.~\ref{sec:kerr}, is the driven-dissipative Kerr resonator, a canonical model of nonlinear quantum optics, displaying complex phenomena such as optical bistability and multiphoton resonances~\cite{DrummondWalls1980,Bartolo2016,Casteels2017}, and also commonly used to describe anharmonic cQED devices such as transmons and Josephson parametric amplifiers~\cite{blais_circuit_2021}. 
We study the dynamics under the influence of a second drive tone or correlated detuning noise. 
In the bistable regime, we find that correctly accounting for the spectral density of the perturbation is essential to predict the switching rate between the bright and dim branches, an effect that Markovian noise models miss.

The second example, described in Sec.~\ref{sec:rc}, is a repetition code with $N=11$ qubits, a standard model of continuous quantum error correction~\cite{paz_zurek_1998}, investigated in the presence of colored frequency noise and coherent errors. In such systems, the computational subspace is, by construction, only weakly perturbed by noise and well separated from rapidly decaying modes, making them natural applications for LSS. This separation of timescales allows LSS to capture the dynamics using a reduced basis of dimension $d_{\rm rb}=37$, compared with a full Liouville-space dimension of $4^{11}\approx 4.2\times10^6$.

In Sec.~\ref{sec:comparison}, we compare LSS to other numerical and analytic approaches to simulate time-dependent dynamics in driven open quantum systems. 
The appendices give the pseudo-code, error estimations, a derivation of the second-order Born approximation, and the numerical propagation scheme used throughout.

\section{The Liouvillian spectral snapshot method}
\label{sec:method}

Consider a time-dependent open quantum system in which the density matrix evolves as
\begin{equation}
    \partial_t\rho(t) = \mathcal L(t)\rho(t).
\end{equation}
Here, $\mathcal L(t) = \mathcal L_0 +\mathcal V(t)$ is the time-dependent Liouvillian, with $\mathcal L_0$ the static Liouvillian and $\mathcal V(t)$ a time-dependent perturbation that can arise from deterministic contributions (e.g., periodic drives) or stochastic fluctuations~\cite{breuerpetruccione2007}. 
The corresponding completely positive trace preserving (CPTP) propagator from time $t=0$ to $t=T$ can be written as the time-ordered exponential~\cite{Rivas_2012}
\begin{equation}\label{eq:cptp_prop}
    \mathcal U_{T,0} = \mathcal T\mathrm{e}^{\int_0^T\!\!\dd t\mathcal L(t)}.
\end{equation}
Computing $\mathcal U_{T,0}$ requires evolving a complete basis of the Liouville space, which is computationally demanding even for moderate-dimensional Hilbert spaces.
For noisy systems, this evolution must be repeated over many realizations to obtain the noise-averaged propagator, further increasing the computational cost.

The processes that remain observable on the scale of the evolution time $T$ are encoded in the  eigenvalues $\upsilon_n$ and corresponding eigenoperators of $\mathcal U_{T,0}$ with $|\upsilon_n|\approx 1$. These slow modes capture, for example, how time-dependent perturbations modify the steady state or the switching rate between metastable states. Processes decaying on time scales much shorter than $T$ appear in the propagator only indirectly, through how they renormalize the slow ones, which is what the reduced basis is designed to capture.

\subsection{Approximating the propagator}

\begin{figure*}[t]
  \centering
  \includegraphics[width=\textwidth]{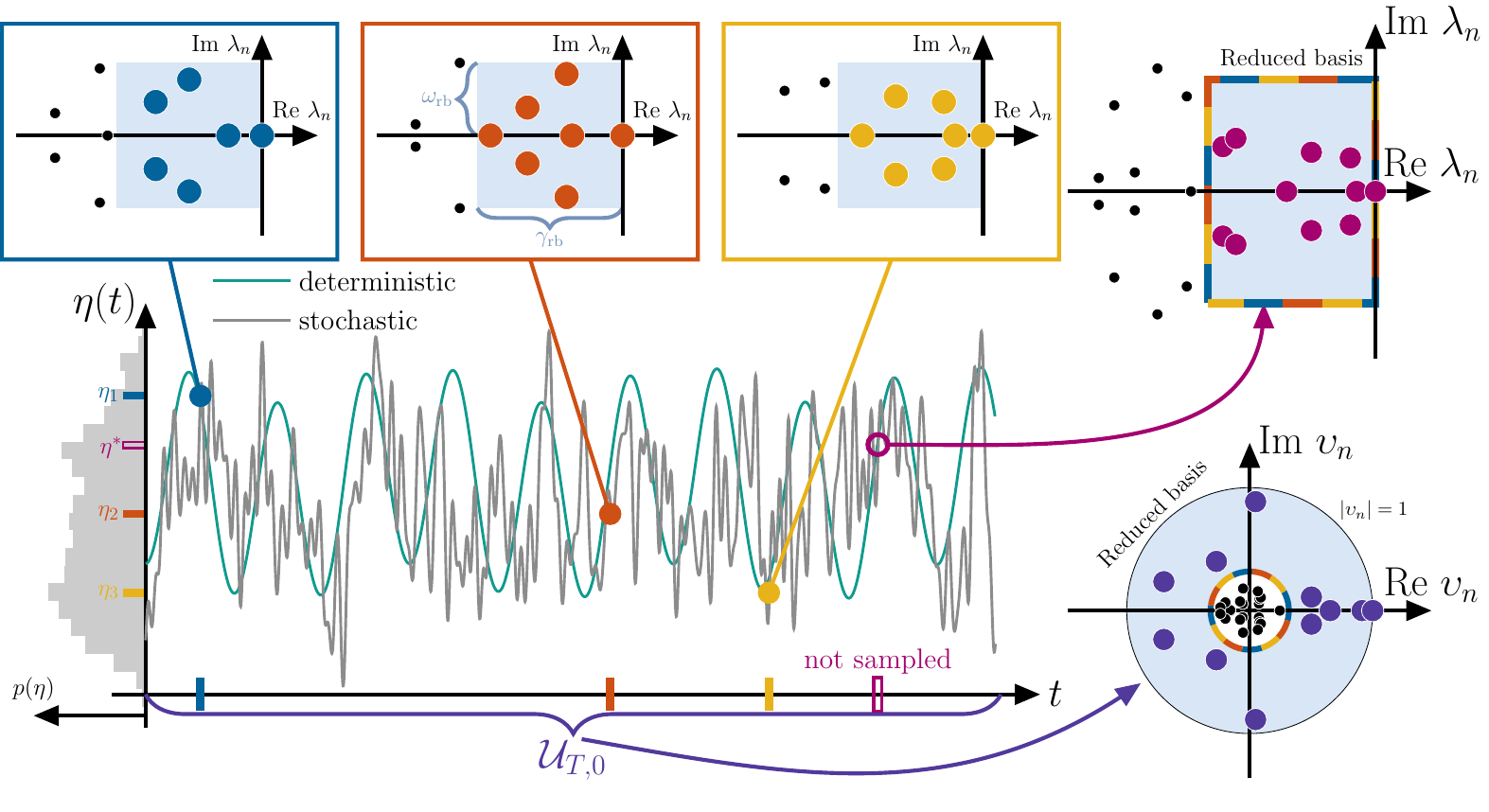}
  \caption{
  \textbf{Illustration of the Liouvillian spectral snapshot (LSS) method.}
    We consider a time-dependent Liouvillian $\LL(t)=\LL_0+\mathcal V(t)$ combining a static generator $\LL_0$ with a perturbation $\mathcal V(t)$, illustrated here for the separable case $\mathcal V(t)=\eta(t)\mathcal V$.
    The perturbation may represent a deterministic drive or colored noise, depicted schematically by the grey and teal curves.
    To cover the range of configurations explored by the perturbation, we select representative Liouvillians by sampling the distribution $p(\eta)$ (left). 
    For each sampled value $\eta_j$, we hold the perturbation fixed and compute the slow eigenoperators of the resulting static Liouvillian $\LL_j=\LL_0+\eta_j\mathcal V$.
    These eigenoperators form a \emph{spectral snapshot}, describing the slow dynamics at that fixed perturbation value.
    The three sampled configurations $\LL_1$, $\LL_2$, and $\LL_3$ are indicated by blue, orange, and gold filled markers, with matching ticks on the schematic time traces.
    Their spectra are shown in the top panels, with eigenvalues satisfying $-\mathrm{Re}\,\lambda_n<\gamma_{\rm rb}$ and $|\mathrm{Im}\,\lambda_n|<\omega_{\rm rb}$ shown as large colored dots, while those outside this window are depicted as small black dots.
    We combine the retained eigenoperators from all snapshots and use a singular-value decomposition to remove redundant directions, obtaining a common reduced basis and its projector $\mathcal P^{\rm rb}$.
   This basis generalizes across the perturbation range, representing the slow eigenoperators at new values of $\eta$ within the same reduced subspace.
   The top-right panel illustrates this for $\LL^*=\LL(t^*)$ (magenta, open symbols).
    We then project the time-dependent master equation onto the same fixed basis to approximate the propagator $\mathcal U_{T,0}$ over the interval $[0,T]$ (bottom right).
    The reduced propagator targets the slowly decaying modes, whose eigenvalues $\upsilon_n$ lie near the unit circle, while fast-decaying modes lie closer to the origin.}
  \label{fig:scheme}
\end{figure*}

The aim of LSS is to construct a compact reduced basis (rb) that captures the slow dynamics while reducing the computational cost of its simulation.
Let $\mathcal P^{\rm rb}$ denote the orthogonal projector onto the subspace spanned by this basis with the instantaneous projected Liouvillian given by $\mathcal L^{\rm rb}(t) = \mathcal P^{\rm rb}\mathcal L(t) \mathcal P^{\rm rb}$.
The basis is chosen so that the corresponding reduced propagator,
\begin{equation}\label{eq:definition_reduced_propagator}
     \mathcal{U}_{T, 0}^{\rm LSS} \equiv \mathcal T \mathrm{e}^{\int _0^T\!\!\dd t' \mathcal L^{\rm rb}(t')},
\end{equation} 
faithfully approximates the slowly decaying eigenvalues $|\upsilon_n| \simeq 1$ and corresponding eigenoperators of the full propagator $\mathcal{U}_{T, 0}$.

For a time-independent Liouvillian, $\mathcal{L}(t) =\mathcal L$, the slowly decaying eigenoperators of $\LL$ are a natural choice for the reduced basis.
Introducing the eigenvalues $\lambda_n$, right eigenoperators $\rho_n$, and left eigenoperators $\sigma_n$~\cite{macieszczak2016} we have
\begin{equation} \label{eq:definition_spectrum}
\begin{split}
    \LL \ket{\rho_n} = \lambda_n \ket{\rho_n}, &\,\,\, \,\, \bra{\sigma_n} \LL = \bra{\sigma_n} \lambda_n,  \\ \braket{\sigma_m}{\rho_n} &=\delta_{m,n}
\end{split}    
\end{equation}
where $\ket\rho$ denotes the vectorized operator $\rho$ and $\braket{\sigma}{\rho}=\mathrm{Tr}[\sigma^\dag \rho]$. The real and imaginary parts of $\lambda_n$ determine the decay rate and oscillation frequency of the mode $n$, respectively.\footnote{Since 
all $\lambda_n$ have non-positive real part, we follow the standard convention in the literature and order the eigenvalues so that $\mathrm{Re}\,\lambda_0\geq\mathrm{Re}\,\lambda_1\geq\mathrm{Re}\,\lambda_2\dots$, with $\lambda_0=0$ corresponding to the steady state $\rho_0\propto\rho_{\rm ss}$.}
We construct the reduced basis by retaining modes with decay rates below $\gamma_{\rm rb}$ and oscillation frequencies below $\omega_{\rm rb}$, namely
\begin{equation}
-\mathrm{Re}\,\lambda_n<\gamma_{\rm rb},
\qquad
|\mathrm{Im}\,\lambda_n|<\omega_{\rm rb}.
\end{equation}
These cutoffs define the retained region of the complex spectrum, characterized by $\lambda_{\rm rb}=-\gamma_{\rm rb}+i\omega_{\rm rb}$ and illustrated in Fig.~\ref{fig:scheme}. To extend this construction to time-dependent dynamics, first consider a Liouvillian that switches at time $\tau$ from $\LL_0 \equiv \LL(t<\tau)$ (i.e. $\mathcal V(t<\tau) =0$) to $\LL_1 \equiv \LL(t>\tau)  =\LL_0+\mathcal V$.
The full propagator reads
\begin{equation}
    \mathcal U_{T,0} = \mathcal U_{T,\tau}\mathcal U_{\tau,0}=\mathrm{e}^{\mathcal L_1(T-\tau)}\mathrm{e}^{\mathcal L_0 \tau}.
\end{equation}
For sufficiently weak perturbations, the slow subspaces of $\mathcal L_0$ and $\mathcal L_1$ overlap strongly, so the reduced basis of $\mathcal L_0$ needs to be enlarged only by the directions of the slow subspace of $\mathcal L_1$ that are not already represented.
This combined basis defines a single projector $\mathcal P^{\rm rb}$ that describes the slow dynamics before and after the switch.
The same argument applies to a general time-dependent perturbation. In Trotterized form, $\mathcal U_{T,0} = \prod e^{\LL(t) \Delta t}$, each factor is generated by a fixed Liouvillian, so a common basis assembled from the slow eigenoperators at representative snapshots describes the whole evolution. This presumes that the time-dependence varies slowly enough that it does not induce leakage into the discarded sector, as we will discuss below. 

\subsection{Greedy construction of the projector}
\label{sec:algorithm}   

Computing the slow eigenoperators of $\mathcal L(t)$ at every time step would require many full-space diagonalizations, each adding little information when the instantaneous slow subspaces overlap strongly.
LSS instead samples the Liouvillian at a few representative values of the perturbation and computes its eigenoperators within the spectral window set by $\lambda_{\rm rb}$.
These \textit{spectral snapshots} are then combined into a common reduced basis, as illustrated in Fig.~\ref{fig:scheme}.
We collect the vectorized eigenoperators as columns of a snapshot matrix and perform a singular-value decomposition (SVD) to remove redundant directions while retaining a compressed description of the combined slow subspace~\cite{BennerGugercinWillcox2015}.
The resulting basis defines the projector $\mathcal P^{\rm rb}$ used in Eq.~\eqref{eq:definition_reduced_propagator}.

The number and the perturbation values at which the spectral snapshots are taken are determined adaptively through a greedy strategy~\cite{Haasdonk2013,Herbst2022,Brehmer2023}. 
At each iteration, we draw a candidate perturbation value and assess whether the current reduced basis adequately represents the slow eigenoperators of the corresponding Liouvillian $\mathcal L(t)$.
To do so, we diagonalize the projected generator $\mathcal L^{\rm rb}(t)=\mathcal P^{\rm rb}\mathcal L(t)\mathcal P^{\rm rb}$ within the reduced subspace, obtaining the eigenpairs $(\tilde\lambda_n,\ket{\tilde\rho_n})$.
We then check how closely these satisfy the eigenvalue equation of the full Liouvillian by evaluating the residual
\footnote{For dense spectra, we apply  the residual test only to reduced eigenpairs inside a smaller validation window, $-\mathrm{Re}\,\tilde\lambda_n<\gamma_{\rm valid}$ and $|\mathrm{Im}\,\tilde\lambda_n|\leq\omega_{\rm valid}$, with $\gamma_{\rm valid}\leq\gamma_{\rm rb}$ and $\omega_{\rm valid}\leq\omega_{\rm rb}$.
As the perturbation varies, eigenvalues near the boundary of the snapshot window can cross it, leaving the corresponding modes represented in some snapshots but absent from others.
Restricting validation to the inner window avoids requiring these boundary modes to converge at every sampled value.}
\begin{equation}\label{eq:residual}
r_n=\bigl\lVert(\mathcal L(t)-\tilde\lambda_n)\ket{\tilde\rho_n}\bigr\rVert,
\end{equation}
where $\ket{\tilde\rho_n}$ is normalized and expressed in the full Liouville space.
If $r_n<\epsilon_{\rm res}$ for all tested eigenpairs, every slow mode represented in the current basis is converged~\footnote{This test cannot detect a slow mode of $\LL(t)$ with negligible overlap with the basis. A cheap post-hoc residual check at $\eta=\pm3\sigma$, or a count of in-window eigenvalues, can further guard against this.}.
Otherwise, we enrich the basis by computing the slow eigenoperators of $\mathcal L(t)$ within the spectral window set by $\lambda_{\rm rb}$ and appending them to the previous basis.
We then perform an SVD compression to remove redundant directions before proceeding to the next sample~\footnote{We discard directions whose singular values are smaller than a fraction $\epsilon_{\rm SVD}$ of the largest singular value.
After the compression step, we check whether the dimension of $\mathcal P^{\rm rb}$ has increased. If it has not, the enriched directions were discarded by the truncation, meaning $\epsilon_{\rm SVD}$ is too permissive. 
We then reduce $\epsilon_{\rm SVD}\to\epsilon_{\rm SVD}/c$, with $c$ of order ten, and repeat the compression. This makes $\epsilon_{\rm SVD}$ adaptive and dependent on $\epsilon_{\rm res}$.}.
The loop terminates once $m_{\rm stop}$ consecutive samples pass the residual test.
Requiring several consecutive successes rather than a single favorable draw prevents the algorithm from terminating the search prematurely.
Screening each candidate requires only a reduced-space eigenvalue problem and sparse matrix--vector products with the full Liouvillian, allowing many candidates to be tested without full-space diagonalization.
For stochastic noise, we draw the snapshot candidates from a distribution $p(\eta)$ rather than sampling it along a single trajectory. This is particularly important for slowly varying noise, where any one realization explores only a narrow portion of $p(\eta)$ within the simulated window. 
Sampling across the distribution helps capture large excursions that may dominate rare-event statistics.
We impose two additional constraints on the reduced basis. To enforce exact trace preservation, we always include the vectorized identity matrix in the reduced basis.
We also retain eigenoperators together with their Hermitian conjugates, so that $\mathcal P^{\rm rb}$ commutes with Hermitian conjugation and the reduced time-evolution preserves Hermiticity. The complete basis-construction procedure is laid out in detail in Algorithm~\ref{alg:lss}.

The reduced dimension makes direct integration of the projected equation of motion computationally feasible, and we adopt this approach throughout this work, as explained in Algorithm~\ref{alg:propagator}. Instead of direct integration, the LSS method could also be combined with other integration methods such as the Magnus expansion~\cite{blanes2009}.

\subsection{Leakage to discarded sector}
Once the LSS algorithm has converged, the reduced basis is expected to faithfully represent the slow eigenoperators of the instantaneous Liouvillian throughout the interval $0\leq t\leq T$.
However, this does not by itself ensure that the reduced propagator captures the full evolution.
Time-dependent perturbations can also couple the retained subspace to the discarded sector $\mathds 1-\mathcal P^{\rm rb}$, introducing errors.
A simple first-order perturbation argument provides a qualitative criterion for the validity of LSS in this regard (see App.~\ref{app:errors} for a more quantitative analysis).
Decomposing a perturbation that is periodic with frequency $\omega$ as $\mathcal V(t)=\sum_m\mathcal V_m\mathrm{e}^{\mathrm im\omega t}$, each component $\mathcal V_m$ transfers weight from a retained mode $k$ to a discarded mode $n$ of $\mathcal L_0$ with an amplitude given by
\begin{equation}\label{eq:resolvent}
    \varepsilon_{\mathrm{leak},m}^{k\to n} =
    \frac{\bigl|\mel{\sigma_n}{\mathcal V_m}{\rho_k}\bigr|}
         {\bigl|\lambda_k-\lambda_n+\mathrm im\omega\bigr|}.
\end{equation}
Thus, excitations into the discarded modes are enhanced near the resonance condition $m\omega\simeq\operatorname{Im}(\lambda_n-\lambda_k)$, where the denominator reduces to $|\mathrm{Re}\,\lambda_n-\mathrm{Re}\,\lambda_k|$. For modes discarded by the decay cutoff, this is at least $\gamma_{\rm rb}-|\mathrm{Re}\,\lambda_k|$, the distance of the retained mode from the edge of the window. The reduced basis therefore remains valid as long as the coupling is weak compared with this distance, meaning that $\gamma_{\rm rb}$ should be placed in a gap in the spectrum.

In closed quantum systems, leakage can only be suppressed by cancellation of fast oscillating terms. In open quantum systems, dissipation also limits leakage because weight transferred to modes beyond the decay cutoff decays at a rate of at least $\gamma_{\rm rb}$. The leaked weight therefore saturates at a level set by the balance between excitation and decay, instead of accumulating over the evolution. Modes that are discarded by the frequency cutoff may decay slower so $\omega_{\rm rb}$ must exceed all frequencies at which the perturbation has significant spectral weight.

\section{Example 1: Driven-dissipative Kerr resonator}
\label{sec:kerr}

\begin{figure*}[th!]
    \centering
    \includegraphics[width=\linewidth]{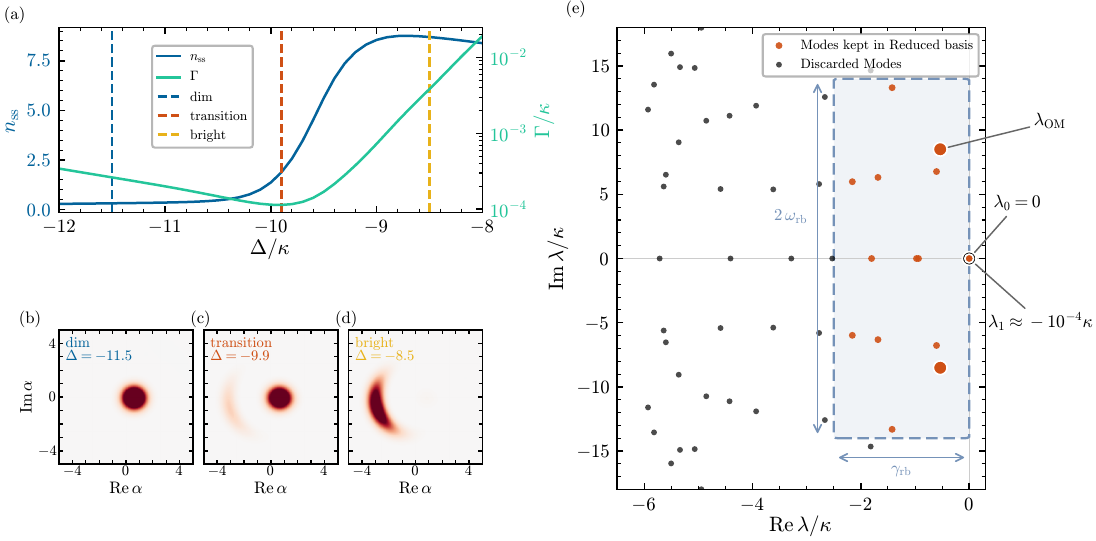}
    \caption{\textbf{Properties of the unperturbed time-independent Kerr resonator.}
    (a) Steady-state photon number, $n_{\rm ss}$ (blue) and switching rate $\Gamma=-\mathrm{Re}\,\lambda_1$ (green) vs the detuning $\Delta$. Around the minimum of the switching rate $\Delta\approx-9.9\kappa$, a sharp increase in photon number signals the passage from a dim to a bright phase.
    (b-d) Steady-state Wigner function at the dim ($\Delta=-11.5\kappa$), transition ($\Delta=-9.9\kappa$) and bright $(\Delta=-8.5\kappa$) points.
    (e) The Liouvillian spectrum at $\Delta = -9.9\kappa$, where the Liouvillian gap is minimal.
    The boxed region indicates the eigenvalues and eigenoperators that can be selected for the reduced basis construction.
    Namely, having fixed the cutoff, $\lambda_{\rm rb}=i\omega_{\rm rb}-\gamma_{\rm rb}$, the eigenvalues that lie in the complex window
    $[-\gamma_{\rm rb},0]+i[-\omega_{\rm rb},\omega_{\rm rb}]$ are selected while those that lie outside are discarded. The window shown is $\gamma_{\rm rb}=2.5\kappa$ and $\omega_{\rm rb}=14\kappa$.
    We mark the steady state $\lambda_0 =0$, the Liouvillian gap $\lambda_1$, and the eigenvalue of the longest-lived oscillating mode, $\lambda_{\rm OM}$.
    Parameters: $F=6.2\kappa$, $K = 0.625\kappa$.}
     \label{fig:kerr_spectrum}
\end{figure*}

\begin{figure}[!tbh]
    \centering
    \includegraphics[width=\linewidth]{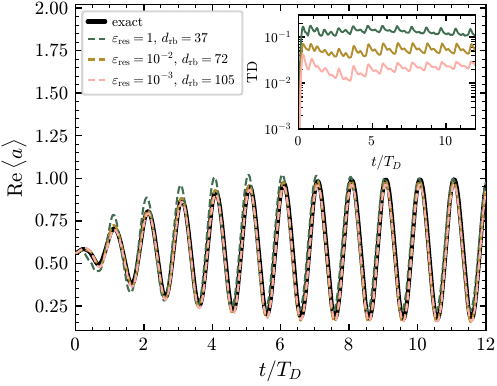}
    \caption{\textbf{Driven dynamics computed in the LSS reduced subspace.} Real part of $\langle a\rangle$ for the Kerr resonator at  $\Delta = -11.5\kappa$, under a second periodic drive with detuning $\Delta_D = -10\,\kappa$ and amplitude $\varepsilon_D = 6\,\kappa$, starting from the steady state of $\mathcal L_0$. The exact
solution of the master equation (black) is compared with the reduced dynamics on three
bases built by the greedy LSS algorithm with $\gamma_{\rm rb}=5\kappa$, $\omega_{\rm rb}=25\kappa$, and
residual tolerances $\varepsilon_{\rm res} = 1$, $10^{-2}$ and $10^{-3}$, giving bases of dimension 37, 72 and 105. Inset: Trace distance of the density matrix computed with the exact and the LSS propagator.}
    \label{fig:kerr_driven_trajectories}
\end{figure}

\begin{figure}[!tbh]
    \centering
    \includegraphics[width=\linewidth]{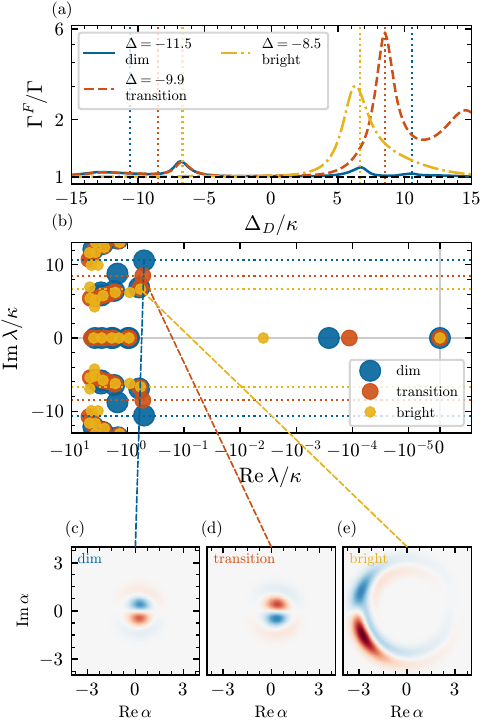}
    \caption{\textbf{Switching rate under a periodic drive.} LSS bases built at the three detunings $\Delta/\kappa=-11.5,-9.9,-8.5$ with $\gamma_{\rm rb}=5\kappa$, $\omega_{\rm rb}=25\kappa$, and $\epsilon_{\rm res}=10^{-3}$ giving LSS basis dimensions $d_{\rm rb} = 105/97/103$. \textbf{(a)} Slowest Floquet decay rate,  $\Gamma^F$, of the reduced one-period propagator, normalized to the static switching rate, versus drive detuning $\Delta_D$ with $\varepsilon_D=0.1\kappa$. Dotted verticals mark $\pm\operatorname{Im}\lambda_{\rm OM}$ at each detuning, with OM the longest-lived oscillating mode. \textbf{(b)} Static Liouvillian spectra at the three detunings (symmetric-logarithmic real axis), with the OM lines (a) drawn horizontally. \textbf{(c--e)} Wigner representation of the OM right eigenoperators.}
  
    \label{fig:floquet}
\end{figure}

\begin{figure}[!tbh]
    \centering
     \includegraphics[width=\linewidth]{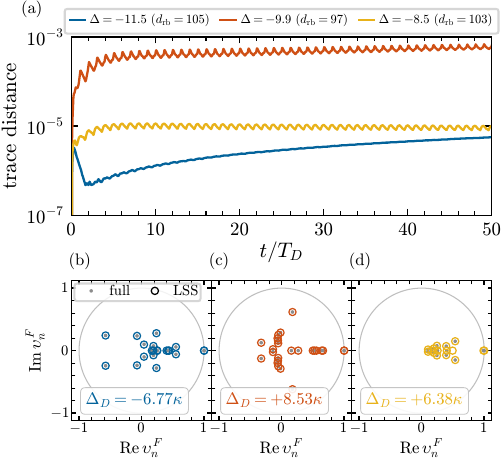}
    \caption{\textbf{Floquet-propagator benchmark of the LSS method.} Same LSS basis as in Fig.~\ref{fig:floquet} at the three detunings. The drive detuning, $\Delta_D$ is set by the peak switching rate according to Fig.~\ref{fig:floquet}.
  \textbf{(a)} Trace distance over 50 drive periods between the exact state and the reduced state, both started from the undriven steady state.
  \textbf{(b--d)} Spectrum of the one-period propagator $\mathcal U_F$: eigenvalues $\upsilon^F_n$ of the full propagator (filled dots) and the LSS propagator (open circles). Only eigenvalues with $\abs{\upsilon_n^F}>0.1$ are shown. 
  The Floquet rate, $\Gamma^F$, agrees with the full propagator to  $+0.4\%/{-1.5\%}/{+0.3\%}$.}
    \label{fig:benchmark}
\end{figure}

We consider a resonator with Kerr non-linearity $K$, single-photon drive with amplitude $F$ and angular frequency $\omega_p$, and single-photon loss with rate $\kappa$.
In the rotating frame of the drive the Liouvillian reads~\cite{DrummondWalls1980, Bartolo2016, RobertsClerk2020}
\begin{equation}\label{Eq:LME_Kerr}
    \mathcal L_0 \rho=-i[\hat H_{\rm Kerr}, \, \cdot\,]\rho+\kappa\mathcal D_{\hat a }\rho,
\end{equation}
where
\begin{equation}
    \hat H_{\rm Kerr} = \Delta \hat a^\dag \hat a +K\hat a^\dag \hat a^\dag \hat a \hat a + F\hat a^\dag + F^*\hat a.
\end{equation}
Here, $\Delta=\omega_0-\omega_p$ is the detuning of the pump with respect to the bare resonator frequency $\omega_0$, and the dissipator is defined by $\mathcal D_{\hat L}\rho=\hat L\rho \hat L^\dag-\tfrac12\{\hat L^\dag \hat L,\rho\}$~\cite{breuerpetruccione2007}, with $\hat{L}=\hat{a}$ describing single-photon loss.

In the thermodynamic limit of large photon number ($F\to\infty$ and $K\to0$ at fixed $F^2K$), the Liouvillian gap closes ($\Gamma\to 0 $) at a critical detuning, signaling a first-order dissipative phase transition~\cite{Bartolo2016, Minganti2018, Casteels2017}.
Close to this bistability, there is a finite fluctuation-activated rate of switching between the two metastable branches that coincides with the Liouvillian spectral gap $\Gamma=-\mathrm{Re}\,\lambda_1$, which we therefore refer to as the switching rate in the following~\cite{Andersen2020} \footnote{$\Gamma$ indicates the bare switching rate for a static Liouvillian ($\mathcal{V}(t) = 0$), while $\Gamma^F$ and $\na{\Gamma}$ denote Floquet or noise-averaged rates, respectively.}.
In Fig.~\ref{fig:kerr_spectrum}, we show some properties of the system across this transition, with fixed parameters $F=6.2\kappa$ and $K=0.625\kappa$. 
In Fig.~\ref{fig:kerr_spectrum}(a) we show the steady-state photon number $n_{\rm ss} = \Tr{\hat a^\dag \hat a \rho_{\rm ss}}$ and the switching rate, $\Gamma=-\mathrm{Re}\,\lambda_1$, versus the detuning $\Delta$.
The photon number increases sharply as the detuning crosses the transition point $\Delta_* = -9.9 \kappa$. Below the transition $(\Delta<\Delta_*)$ the steady state is a low-photon (dim) state and above it is a high-photon (bright) state [Fig.~\ref{fig:kerr_spectrum}(b,d)]. Near $\Delta_*$ it is a mixture of the two branches and the Wigner function is bimodal [Fig.~\ref{fig:kerr_spectrum}(c)]. 
In Fig.~\ref{fig:kerr_spectrum}(e) we show the Liouvillian spectrum at $\Delta=\Delta_*$. In addition to the steady-state and gap eigenvalues, $\lambda_0=0$ and $\lambda_1$, we highlight $\lambda_{\rm OM}$, corresponding to the longest-lived mode with finite oscillation frequency, ${\rm Im}\,\lambda_{\rm OM}\neq0$~\cite{Haga2024Sep}. 
We refer to it as the oscillating mode (OM) and, as shown below, it dominates the response to time-dependent perturbations.  

\subsection{Two-tone driving}
\label{sec:twotonedrive}

As a first application of LSS, we study the resonator under an additional drive, modeled as the perturbation $\mathcal V(t) =-i[H_D(t),\cdot]$ with
\begin{equation}
    H_D(t) = \varepsilon_D \hat a \mathrm{e}^{-i\Delta_D t}+\varepsilon^*_D\hat a^\dag \mathrm{e}^{i\Delta_D t} ,
\end{equation}
where $\Delta_D = \omega_p - \omega_D$ is the drive detuning and $\varepsilon_D$ its amplitude.
To assess whether the LSS method accurately captures the dynamics under the influence of this extra drive, in Fig.~\ref{fig:kerr_driven_trajectories}, we compare the exact time evolution with LSS dynamics obtained using three bases constructed with different residual tolerances $\varepsilon_{\rm res}$. The bases are constructed with spectral cutoffs $\gamma_{\rm rb}=5\kappa$ and $\omega_{\rm rb}=25\kappa$, $m_{\rm stop}=10$ and $\varepsilon_{\rm res}=1,10^{-2},10^{-3}$. We use $\Delta=-11.5\kappa$, drive detuning $\Delta_D=-10\kappa$ and a strong drive amplitude $\varepsilon_D=6\kappa$. We compute $\langle \hat a \rangle(t) = \Tr{\hat a \rho(t)}$ using either the exact evolution, $\rho_{\rm ex}(t)=\mathcal U_{t,0}\rho_0$, or the reduced-basis approximation $\rho_{\rm rb}(t)=\mathcal U_{t,0}^{\mathrm{LSS}}\rho_0$, starting from the steady state $\rho_0$ of the static Liouvillian $\LL_0$.
When $\varepsilon_{\rm res}=1$, the algorithm stops after computing the first spectral snapshot, which we choose to be $\mathcal L_0$. This gives a reduced subspace of dimension $d_{\rm rb}=37$, resulting in visible deviations in $\expval{a}$ from the exact solution, as seen in Fig.~\ref{fig:kerr_driven_trajectories}. For $\varepsilon_{\rm res}=10^{-2}$ and $10^{-3}$ the LSS algorithm stops after two or three snapshots, respectively, resulting in dimensions $72$ and $105$ with noticeably better accuracy. The inset shows the trace distance, $\mathrm{TD}(t)=\frac{1}{2}\|\rho_{\rm ex}(t)-\rho_{\rm rb}(t)\|_1$ \footnote{Using the trace norm $\|X\|_1=\tr\sqrt{X^\dag X}$}, between the reduced and exact states.

Having tested LSS in the strong drive regime, we now turn to a weak probe, $\varepsilon_D=0.1\kappa$, and vary its detuning. When the probe is resonant with long-lived oscillating modes~\cite{devoret1984}, we expect it to strongly enhance the switching rate. This detuning scan then amounts to two-tone spectroscopy of the switching rate, and is analogous to experiments used to probe the level structure of Kerr parametric oscillators~\cite{yamaji_2022}.

The Liouvillian $\mathcal L(t) = -i[H_{\mathrm{Kerr}}+H_D(t),\cdot]+\kappa\mathcal D_{\hat a}$ is periodic with period $T_D = 2\pi/|\Delta_D|$.
Therefore, we can construct the Floquet propagator $\mathcal U_F \equiv \mathcal U_{T_D,0}=\mathcal T\exp{\int_0^{T_D}\dd t \mathcal L(t)}$, which describes the evolution over one period.
The switching rate is then $\Gamma^F=-\log|\upsilon_1^F|/T_D$ where $\upsilon_1^F$ is the eigenvalue of $\mathcal U_F$ with the largest modulus below unity~\cite{schnell2020}. In the example we study here, $\Gamma^F\ll|\Delta_D|$ meaning that $\Gamma^F$ represents the switching rate of the coarse-grained stroboscopic evolution obtained by repeatedly applying $\mathcal U_F$. 

We construct the LSS basis at the transition point $\Delta=-9.9\kappa$ and in the bright phase $\Delta=-8.5\kappa$ using the same LSS parameters as for $\Delta=-11.5\kappa$ with $\varepsilon_{\rm res}=10^{-3}$, resulting in dimensions 97 and 103, respectively. 
The reduced dimension is about 10\% of the full Liouville-space dimension of $32^2=1024$, where 32 is the Fock-space cutoff required for numerical convergence.
Using the LSS bases, we compute the Floquet propagator and extract the driven relaxation rate $\Gamma^F$. 
In Fig.~\ref{fig:floquet}(a), we plot $\Gamma^F/\Gamma$ versus $\Delta_D$, and observe enhanced switching rates at specific frequencies.
At the transition point, $\Gamma^F$ peaks at $\Delta_D\approx \mathrm{Im}\,\lambda_{\rm OM}$.
The Wigner representations of these eigenmodes (Fig.~\ref{fig:floquet}(c-e)) correspond to excitations localized around the phase-space minima that host the metastable states.
This is a driven-dissipative analog of mode-selective chemistry, where exciting a specific vibration of a reactant changes the barrier-crossing rate~\cite{zare1998}. 
The same effect has been observed in current-biased Josephson junctions, where a weak resonant microwave tone enhances the escape from the zero-voltage state~\cite{devoret1984}. 

In Fig.~\ref{fig:benchmark}, we stress-test the method by comparing the reduced-subspace and exact evolutions at the points with the strongest enhancement of the switching rate, as found in Fig.~\ref{fig:floquet}(a). These points are particularly challenging to capture within the reduced basis because the drive fulfills the resonance condition enhancing the transition amplitude in  Eq.~\eqref{eq:resolvent}, and thus maximizes leakage into the discarded sector. In Fig.~\ref{fig:benchmark}(a) we show the TD for the three detunings. 
The TD saturates once leakage into the discarded sector is balanced by the decay back to the retained sector due to dissipation. The TD is largest at the transition point where it reaches about $4\times 10^{-4}$ after 8 periods and $7\times 10^{-4}$ after 50. The eigenvalues of the Floquet propagators agree accordingly [Fig.~\ref{fig:benchmark}(b-d)]. Specifically, the relative error in $\Gamma^F$ computed from the LSS and exact propagator is $+0.4\%$, $-1.5\%$, and $+0.3\%$ at the dim, transition, and bright points, with $\Gamma^F=6.7\times 10^{-4}\kappa$ at the transition point. At the bright point, there is clearly one eigenvalue of the LSS propagator with no corresponding eigenvalue of the exact propagator. This spurious mode arises from the inclusion of the identity operator in the reduced basis (Sec.~\ref{sec:method}). 
In the cases studied here the spurious mode is always fast decaying, traceless, and has a very small overlap with the rest of the dynamics so it introduces negligible errors into physical observables. One could remove it by excluding the identity vector from the LSS basis, at the cost of potentially breaking trace preservation.

\subsection{Frequency noise and dephasing}

We now study how fluctuations in the detuning of the resonator, $\Delta(t)=\Delta_*+\eta(t)$, modify the noise-averaged switching rate, $\na{\Gamma}$, at the transition point.
To this end, we introduce the propagator for a noise realization, $\eta(t)$, as
\begin{equation}
    \mathcal U_{t,t'}[\eta] = \mathcal T \mathrm{e}^{\int_{t'}^{t}\dd t''\,\qty(\mathcal L_0-i\eta(t'')[\hat a^\dag \hat a,\,\cdot\,])}.
\end{equation}
We model $\eta(t)$ as a real, Gaussian process with zero mean and power spectral density (PSD) $S(f)$, defined by (see also App.~\ref{app:noiseaveraging})
\begin{equation}\label{eq:psddef}
\langle \eta(t)\,\eta(t') \rangle_S
    = \int_{f_{\rm ir}}^{f_{\rm uv}} \dd f\, S(f)\, \cos\qty(2\pi f[t-t']).
\end{equation}
Here, $\langle\cdot\rangle_S$ denotes an average over noise realizations, and $f_{\rm ir}$ and $f_{\rm uv}$ are the infrared and ultraviolet cutoffs of the noise spectrum\footnote{In practice, we sample $R$ time series of noise realizations 
$\{\eta_1(t), \eta_2(t), \ldots, \eta_R(t)\}$, each approximated by a discrete set of points in time, whose sampling rate is fixed by $f_{\rm uv}$.}. Since we assume Gaussian noise with vanishing mean $\langle \eta(t)\rangle=0$, all higher order correlation functions reduce to products of the two-point function Eq.~\eqref{eq:psddef} by Wick's theorem. 

To find $\na{\Gamma}$, we average the propagator over three batches of $R=160$ noise realizations, each integrated to $T\approx 81/\kappa$, and obtain the averaged propagator $\na{\mathcal U}_{T,0}=\langle \mathcal U_{T,0}[\eta]\rangle_S$. The rate is then given by the eigenvalue of largest modulus below unity, $\na{\upsilon}_1$, as $\na{\Gamma}=-\log|\na\upsilon_1|/T$. See App.~\ref{app:sm-strang} and Algorithm~\ref{alg:propagator} for more details on the numerical scheme used to compute the propagators. 

\begin{figure}[t]
\centering\includegraphics[width=\linewidth]{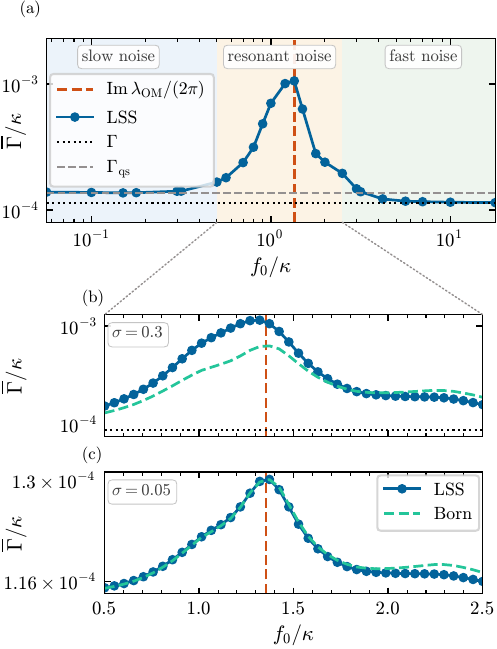}
    \caption{\textbf{Switching rate versus noise center frequency at bistability.}  Noise-averaged switching rate, $\na \Gamma=-\log|\na\upsilon_1|/T$, versus the center frequency $f_0$ of Lorentzian detuning noise with strength $\sigma=0.3\kappa$. The switching eigenvalue $\na\upsilon_1$ is computed  from the LSS propagator with $\gamma_{\rm rb}=2.5\kappa$, $\omega_{\rm rb}=14\kappa$, and $\epsilon_{\rm res}=3\times10^{-5}$ giving a reduced dimension $d_{\rm rb}=99$. 
  \textbf{(a)} Log-scale plot with noise parameters scaled to $f_0$ as $\gamma=f_0/13$, $f_{\rm ir}=10^{-2}\kappa$, and
  $f_{\rm uv}=30f_0$.
  \textbf{(b, c)} Linear zoom in of the resonant band with fixed parameters $\gamma=0.104\kappa$, $f_{\rm ir}=10^{-2}\kappa$, $f_{\rm uv}=45\kappa$, for
  $\sigma=0.3\kappa$ (b) and $\sigma=0.05\kappa$ (c). Dashed teal: second-order Born result (App.~\ref{sec:bornapprox}).}
    \label{fig:kerr_relaxationrate}
\end{figure}

\begin{figure}[t]
\centering
\includegraphics[width=\linewidth]{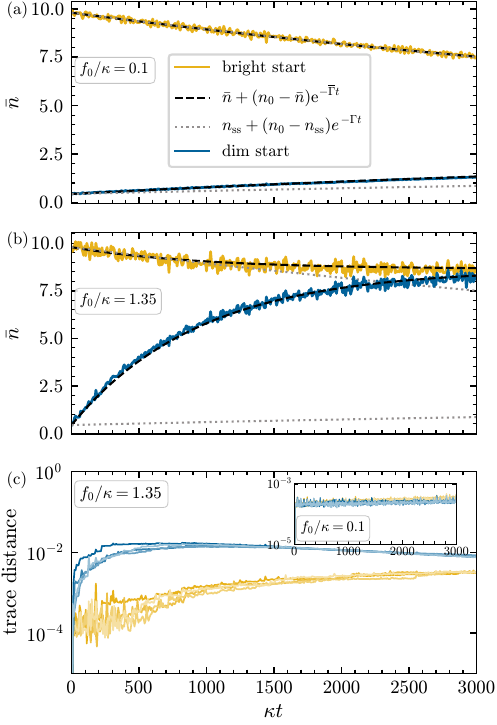}
\caption{\textbf{Branch decay under Lorentzian noise.} Trace distance and photon number evolution computed using the same LSS basis as in Fig.~\ref{fig:kerr_relaxationrate}. \textbf{a}[\textbf{b}] Photon number, averaged over 4 noise realizations, starting from the metastable dim or bright states at $\Delta_*$ with $f_0=0.1\kappa$ [$f_0=1.35\kappa$] and otherwise the same noise parameters as what was used in Fig.~\ref{fig:kerr_relaxationrate}(a). \textbf{c} TD vs time for $f_0=1.35\kappa$ for the 4 trajectories. Inset shows the same at $f_0=0.1\kappa$.}
    \label{fig:kerr_decay}
\end{figure}

Lorentzian noise spectra are known to occur in cQED systems due to coupling to damped modes such as a readout resonator~\cite{gambetta2006}.
The power spectral density for Lorentzian noise reads 
\begin{equation}
\label{eq:psdlorentz}S(f)=\frac{\sigma^2\gamma/\pi}{(f-f_0)^2+\gamma^2}
\end{equation}
with noise variance $\expval{\eta^2}=\int \dd f S(f)=\sigma^2$, width $\gamma$ and central frequency $f_0$. In the time domain, the corresponding trajectories have typical oscillation frequency $f_0$ and correlation time $1/(2\pi\gamma)$. We study how the system responds to noise at different frequencies, $f_0$, and the capability of LSS in capturing these effects. 

To construct the reduced basis, we draw candidate values $\eta_n$ from the Gaussian distribution $p(\eta)=\mathcal N(0,\sigma^2)$ and form the corresponding Liouvillians $\LL_n=\mathcal L_0-i\eta_n[\hat a^\dag\hat a,,\cdot]$. Using $\gamma_{\rm rb}=2.5\kappa$, $\omega_{\rm rb}=14\kappa$, $\epsilon_{\rm res}=3\times10^{-5}$, and $m_{\rm stop}=10$, we obtain a reduced basis of dimension $d_{\rm rb}=99$.

In Fig.~\ref{fig:kerr_relaxationrate}, we show $\na{\Gamma}$ vs $f_0$ and identify three regimes: slow noise, resonant noise, and fast noise.
In the slow noise region, where the noise is slow compared with the relaxation within each metastable branch ($f_0\ll\kappa$) but still fast compared with the switching, the state follows the noise adiabatically within each branch and the switching rate converges to the quasi-static average, $\na{\Gamma}_{\rm qs}\simeq\int \dd \eta \,p(\eta)\Gamma(\Delta_*+\eta)$. 
In the fast noise regime, where $f_0\gg\kappa$, the noise averages out on the timescale of the system. 
Therefore, the slow-time evolution is not affected by the noise and $\na{\Gamma}$ collapses to the noise-free switching rate $\Gamma$. Finally, for resonant noise, the PSD has support at the frequencies of the long-lived oscillating modes.
This induces transitions to long-lived coherences, and as a result, $\na\Gamma$ increases by an order of magnitude compared to $\Gamma$. 
This is an intrinsically dynamical effect that a frequency-independent (white-noise) dephasing dissipator cannot capture.  In Fig.~\ref{fig:kerr_relaxationrate} (b) and (c) we show $\na\Gamma$ vs $f_0$ for $\sigma/\kappa=0.3,0.05$ in the resonant region, and compare with the rate computed from a second-order Born approximation (see App.~\ref{sec:bornapprox} for details). For weak noise, $\sigma/\kappa=0.05$, the numerical LSS computation matches the perturbative prediction. In particular, both methods predict that the maximum switching rate occurs when the resonance condition $2\pi f_0\approx \mathrm{Im}\,\lambda_{\rm OM}$ is met.
Similarly to the detuned drive case discussed in Sec.~\ref{sec:twotonedrive}, this enhancement occurs when noise processes resonantly activate the oscillating mode shown in Fig.~\ref{fig:floquet}(d).
At $\sigma/\kappa=0.3$ the perturbative result significantly underestimates the peak rate, showing that direct numerical simulations are needed to achieve quantitatively accurate predictions. 
We illustrate the distinct features in the slow and resonant regimes in Fig.~\ref{fig:kerr_decay} (a) and (b) by showing the time-evolution, averaged over noise realizations, starting from the bright or dim initial states. 
At $f_0=0.1\kappa$, the evolution stays close to the noise-free one (dotted lines), whereas at resonance, $f_0=1.35\kappa$, the branches relax faster and toward a different steady-state photon number $\na n \approx 8.6$ compared with $n_{\rm ss}\approx 1.9$. This means that the noise primarily enhances escape from the dim branch consistent with the activated OM Wigner function describing oscillations around the dim branch, as seen in Fig.~\ref{fig:floquet}.
The trace distance to the exact evolution is shown in Fig.~\ref{fig:kerr_decay}(c). 
The resulting noise-averaged rate computed in the full Liouville space for the same noise realizations is $\na\Gamma_{\rm ex}=1.020\times 10^{-3}\kappa$ which is $3.4\%$ lower than the LSS rate, $\na \Gamma = 1.056\times10^{-3}\kappa$.   This is a window-limited truncation error at this worst-case point, that decreases with a larger $\gamma_{\rm rb}$, as we will discuss in the following.

\begin{figure}[t]
    \centering
    \includegraphics[width=\linewidth]{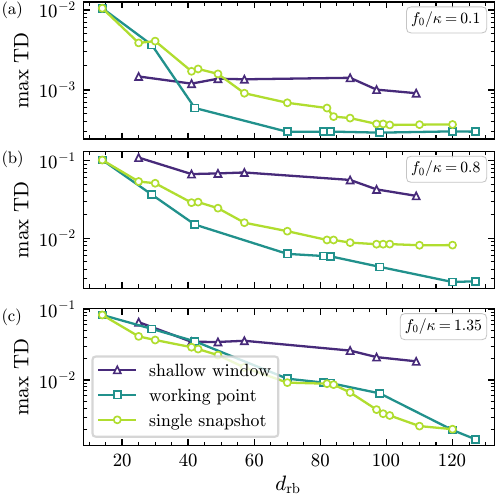}
    \caption{\textbf{Spectral depth versus parameter width in the reduced basis.}
  Maximum trace distance between the exact and the reduced-basis state versus basis
  dimension, $d_{\rm rb}$, for three constructions at $\Delta_*=-9.9\kappa$ under
  Lorentzian detuning noise ($\gamma=f_0/13$, $\sigma=0.3\kappa$) at \textbf{(a)} $f_0=0.1\kappa$ (slow),
  \textbf{(b)} $f_0=0.8\kappa$ (slow part of the resonant regime) and \textbf{(c)} $f_0=1.35\kappa$ (the OM
  resonance). Each point is the maximum over three common noise realizations, with both the dim and bright initial states, and in the time-window $t\kappa\in[0,20]$. Shallow window
  ($\triangle$): $\gamma_{\rm rb}=1.2\kappa$, $\omega_{\rm rb}=14\kappa$, residual tolerance in range
  $\epsilon_{\rm res}\in[10^{-3},10^{-9}]$ to vary $d_{\rm rb}$.
  Working point ($\square$): $\gamma_{\rm rb} =2.5\kappa$, $\omega_{\rm rb}=14\kappa$, and $\epsilon_{\rm res}\in[10^{-1},10^{-6}]$. Single snapshot ($\circ$): one diagonalization at
  $\mathcal L(t)=\mathcal L_0$.}
    \label{fig:kerr_lss_comparison}
\end{figure}

In the previous examples, we reused the same LSS basis to capture the influence of noise with $f_0$ varying over orders of magnitude. However, the optimal LSS parameter choices depend on the characteristic timescale of the perturbation. This is illustrated in Fig.~\ref{fig:kerr_lss_comparison} where we show the maximum trace distance (max TD) versus LSS basis dimension ($d_{\rm rb}$) for three different constructions. In the first two we use the LSS method with either a shallow window, $\gamma_{\rm rb}=1.2\kappa$ or a deeper window with $\gamma_{\rm rb}=2.5\kappa$ as used in Figs.~\ref{fig:kerr_relaxationrate} and \ref{fig:kerr_decay}, and $\omega_{\rm rb}=14\kappa$ in both cases. We vary the resulting dimension by changing $\epsilon_{\rm res}$. In the third construction we take the extreme limit of the deep window where the reduced basis is constructed  from a single snapshot of the unperturbed Liouvillian, $\mathcal L_0$, with the window set at $\omega_{\rm rb}=30\kappa$ and $\gamma_{\rm rb}$ varied to change $d_{\rm rb}$. In Figs.~\ref{fig:kerr_lss_comparison} (a) and (b) where $f_0/\kappa=0.1,0.8$ respectively, the LSS with $\gamma_{\rm rb}=2.5\kappa$ is more accurate than the single snapshot for $d_{\rm rb}\geq40$. At these lower frequencies, it is advantageous to sample several parameter points to better capture the slowly changing Liouvillian. This is in contrast to Fig.~\ref{fig:kerr_lss_comparison}(c) with $f_0/\kappa=1.35$ where the single snapshot basis has the best performance at almost all $d_{\rm rb}$. Here, 
the noise activates modes deep in the spectrum which the single snapshot basis captures better at similar $d_{\rm rb}$, compared to a basis constructed from multiple snapshots. 

\section{Example 2: Repetition code}\label{sec:rc}

\begin{figure}[t]
    \centering
    \includegraphics[width=\linewidth]{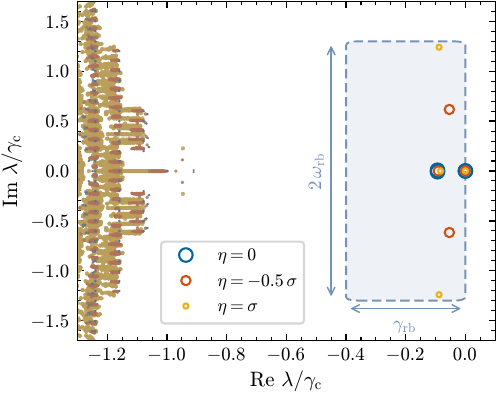}
    \caption{\textbf{Spectrum of the repetition code.} Low-lying spectrum of the instantaneous Liouvillian
    $\mathcal L_k=\mathcal L_0+\eta_k\mathcal V$ for the repetition code, at
    $\eta_k=0$, $-\sigma/2$ and $\sigma$. In all three cases the four
    eigenvalues associated with the logical subspace remain well separated from
    the rest of the spectrum, which lies at $\mathrm{Re}\,\lambda_n\lesssim-0.9\gamma_c$.
    The shaded region illustrates a spectral window separating the four slow modes from the rest. 
    Parameters: $N=7$, $h=0.2\,\gamma_c$, $J=0.05\,\gamma_c$, $\kappa_h=0.04\,\gamma_c$,
    $\sigma=0.1\,\gamma_c$.}
    \label{fig:rc_spectrum}
\end{figure} 

As a second example, we consider a quantum memory made up of $N$ data qubits (with $N$ odd) encoding one logical bit in the repetition code, with code states $\ket{\vec 0}=\ket{0\cdots0}$ and $\ket{\vec 1}=\ket{1\cdots1}$. In a measurement-based
implementation, syndromes are extracted by ancilla qubits that are entangled with the data qubits and measured. 
This procedure measures the stabilizers $\hat Z_j\hat Z_{j+1}$. The code states span their joint $+1$ eigenspace, while states with eigenvalue $-1$ indicate that an error has occurred~\cite{kelly2015,google2025}. Here we study a continuous error-correction model~\cite{paz_zurek_1998,sarovar2005} in which detection and recovery are described by the
engineered dissipator
\begin{equation}\label{eq:pz_dissipator}
  \gamma_c \sum_{b:\,|b|<N/2} \mathcal{D}[\hat R_b],
  \qquad
  \hat R_b = \ket{\vec 0}\!\bra{b} + \ket{\vec 1}\!\bra{\bar b},
\end{equation}
with $\gamma_c$ the correction rate, $b\in\{0,1\}^{N}$ a bit string of weight
$|b|=\sum_j b_j$, and $\bar b$ its complement ($\bar 0 = 1$ and $\bar 1 = 0$). 
Each $\hat R_b$ with
$b\neq\vec 0$ removes the effect of bit flips by returning the two states sharing the syndrome of $b$ to the code
space by majority vote. The $b=\vec 0$ term, instead, only damps coherences between the code space and the remainder of the Hilbert space. 
We monitor the stored bit through
\begin{equation}\label{eq:ZL}
  \hat Z_L=\sum_{b:\,|b|<N/2}\ket{b}\!\bra{b}-\sum_{b:\,|b|>N/2}\ket{b}\!\bra{b},
\end{equation}
which equals $+1$ ($-1$) on every state that majority vote returns to
$\ket{\vec 0}$ ($\ket{\vec 1}$).
This dissipation channel corrects errors in the form of stochastic bit flips, described by the dissipator $\kappa_h\sum_j\mathcal D[\hat X_j]$. 

To demonstrate the versatility of the LSS approach, we go beyond the standard stochastic bit-flip model by including coherent control errors and temporally correlated phase noise, which are common in experimental settings.
Specifically, we consider coherent errors generated by a uniform single-qubit rotation $h\sum_j\hat X_j$ and a uniform two-qubit term $J\sum_j\hat X_j\hat X_{j+1}$ that acts on neighboring pairs, which may emerge from miscalibration of shared controls~\cite{greenbaum2018,bravyi2018}.
The phase noise is modeled as a collective frequency fluctuation  $\eta(t)\sum_j\hat Z_j$ with a $1/f$ spectrum~\cite{paladino_2014}.
Correlated phase noise of this kind has been observed to dominate dephasing in multi-qubit
registers and is highly detrimental to a repetition
code~\cite{clemens2004,monz2011}. 
Note that this type of colored noise cannot be replaced by a dissipator, as it is neither white on the scale of $\gamma_c$, which would be captured by a Markovian rate, nor slow on the scale of the logical dynamics, which would justify a quasi-static average. The resulting model is
\begin{equation}\label{eq:rc_model}
\begin{split}
  \mathcal{L}(t) &=\mathcal{L}_0+\eta(t)\,\mathcal{V},
  \\
  \mathcal{L}_0&=-\mathrm{i}[\hat H_{XX},\cdot\,]
  +\gamma_c\sum_{b:\,|b|<N/2}\mathcal{D}[\hat R_b]
  +\kappa_h\sum_{j=1}^{N}\mathcal{D}[\hat X_j],
  \\
  \mathcal{V}&=-\mathrm{i}\Big[\sum_{j=1}^{N}\hat Z_j,\cdot\,\Big],
\end{split} 
\end{equation}
with
\begin{equation}
  \hat H_{XX}=h\sum_{j=1}^{N}\hat X_j+J\sum_{j=1}^{N-1}\hat X_j\hat X_{j+1}.
\end{equation}
In addition to directly inducing phase flips, the noise term can enhance bit flip processes by modulating the energy differences between states coupled by $\hat H_{XX}$. 
This requires $\eta(t)$ to have spectral weight at frequencies where $\hat H_{XX}$ can introduce errors. 
This is reminiscent of the enhancement of the Kerr switching rate by noise resonant with the oscillating mode [Fig.~\ref{fig:kerr_relaxationrate}], and a strong motivation to include time-dependent processes in the numerical simulation of open systems.

\begin{figure}[t]
    \centering
    \includegraphics[width=\linewidth]{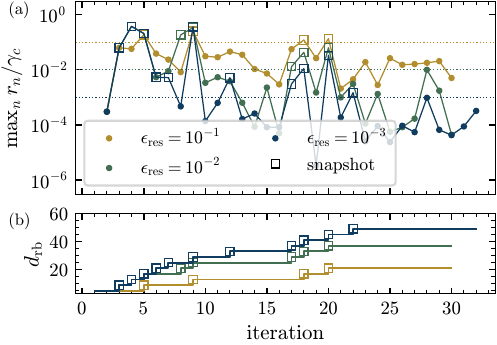}
    \caption{\textbf{Greedy construction of the LSS basis for the repetition code with $N=11$ qubits.}
    (a) Largest residual, Eq.~\eqref{eq:residual}, versus the iteration of the LSS construction. 
    The dotted horizontal lines represent the three values of residual tolerances
    $\epsilon_{\rm res}=10^{-1},10^{-2},10^{-3}$ considered. 
    While a new Liouvillian is sampled at each  iteration, it is only an excursion above the tolerance (empty square marker) that triggers a diagonalization and an enlargement of the basis.
    Sampling stops after $m_{\rm stop}=10$ consecutive draws pass the residual test.
    (b) Corresponding reduced basis dimension $d_{\rm rb}$.
    }
    \label{fig:rc_greedy}
\end{figure}

Here, we consider a pink-noise PSD, $S(f)\sim 1/f$
with variance $\langle\eta^2\rangle=\sigma^2$, $\sigma=0.1\,\gamma_c$, and cutoffs
$f_{\rm ir}=\gamma_c/400$ and $f_{\rm uv}=2\gamma_c$. Throughout we use the static parameters  
$h=0.2\gamma_c$, $J=0.05\gamma_c$ and $\kappa_h=0.04\gamma_c$. All trajectories start from $\ket{\vec 0}$ relaxed for a time $4/\gamma_c$ with $\eta=0$, which removes fast-decaying components, so that $\langle \hat Z_L\rangle(0) \approx 0.88$ in the simulations.

In Fig.~\ref{fig:rc_spectrum} we show the spectrum of $\mathcal L_k$ for
$\eta_k=0$, $-\sigma/2$ and $\sigma$ for a chain of $N=7$ qubits, small enough that the entire spectrum can be computed. In all three cases the four
low-lying eigenvalues, which encode the logical subspace, remain well separated
from the remaining modes, which lie at $\mathrm{Re}\,\lambda_n\lesssim-0.9\gamma_c$.
This clean spectral separation makes the model particularly well suited for
LSS. We use a spectral window $\gamma_{\rm rb}$ and $\omega_{\rm rb}$ such that the four slow eigenmodes are always included while the rest of the spectrum is discarded.  
 
We then consider a $N=11$ qubit chain whose Liouville space has dimension $4^{11}\approx 4.2\times10^6$, constructing the LSS basis from snapshots $\mathcal L_k=\mathcal L_0+\eta_k\mathcal V$,
with $\eta_k$ drawn from the Gaussian distribution $\mathcal N(0,\sigma^2)$.
Figure~\ref{fig:rc_greedy}(a) shows the greedy construction for three residual
tolerances, $\epsilon_{\rm res}=10^{-1}$, $10^{-2}$ and $10^{-3}$. Whenever the
largest residual of Eq.~\eqref{eq:residual} exceeds the tolerance, we diagonalize
$\mathcal L_k$ within the window and enlarge the basis. The sampling stops after
$m_{\rm stop}=10$ consecutive draws pass the residual test (Sec.~\ref{sec:algorithm}).
Figure~\ref{fig:rc_greedy}(b) shows the corresponding growth of $d_{\rm rb}$.

\begin{figure}[t]
    \centering
    \includegraphics[width=\linewidth]{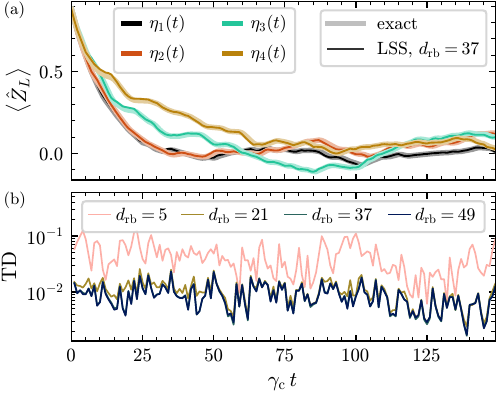}
    \caption{\textbf{Single noise realizations of the repetition code}. (a) Evolution of the
    logical $\langle\hat{Z}_L\rangle$ for four realizations of noise, $\eta_r(t)$,
    computed by exact integration in the full Liouville space (large solid) and with LSS
    (small solid, $d_{\rm rb}=37$). (b) Trace distance between
    the exact and LSS states versus time for the same noise realization, $\eta_1(t)$.}
    \label{fig:rc_trajectory}
\end{figure}

In Fig.~\ref{fig:rc_trajectory}(a) we compute the evolution of the
logical $\langle\hat{Z}_L\rangle$ for several noise realizations, by exact
integration in the full $4^N$-dimensional Liouville space and compare it with LSS in the basis with $d_{\rm rb}=37$ obtained for $\epsilon_{\rm res}=10^{-2}$.
In all examples, LSS agrees closely with the exact dynamics.
In Fig.~\ref{fig:rc_trajectory}(b) we apply the more stringent criterion of the
trace distance to the exact state and show it for different reduced basis dimensions $d_{\rm rb}=5,21,37,49$ for the same noise realization $\eta_1$. This shows that enlarging the basis initially improves the accuracy substantially, but saturates at $d_{\rm rb}=21$. 
This residual error is thus set by leakage into modes outside the spectral window.

\begin{figure}[t]
    \centering
    \includegraphics[width=\linewidth]{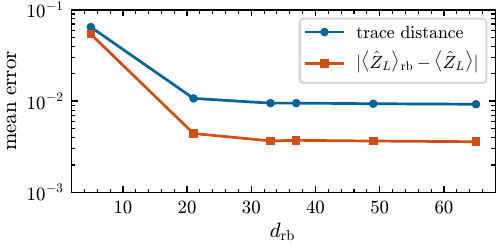}
    \caption{\textbf{Accuracy versus basis dimension.} Mean error averaged over four noise realizations and over $\gamma_ct\in[0,150]$, as a function of
    the basis dimension $d_{\rm rb}$. Trace distance between exact and LSS states
    ($\circ$) and error on the logical $\langle\hat{Z}_L\rangle$ ($\square$).}
    \label{fig:mean_error}
\end{figure}

The saturation with $d_{\rm rb}$ is also seen in Fig.~\ref{fig:mean_error} which shows the trace distance and the error in $\langle Z_L\rangle$ averaged over time and the four noise realizations vs $d_{\rm rb}$. The
error on the logical $\langle\hat{Z}_L\rangle$ is about three times
smaller. This is expected as the trace distance is sensitive to the full state,
including the component that has leaked into the discarded sector, while
$\hat{Z}_L$ is supported on the retained slow modes and is therefore less sensitive to the leakage.

These results suggest further ways to improve and extend LSS.
First, the time evolution error could be further improved by \textit{dynamical enrichment} of the reduced basis. 
Namely, having selected the initial eigenoperators within a certain window $[\gamma_{\rm rb}, \omega_{\rm rb}]$ as in Fig.~\ref{fig:scheme}, one could further enrich the basis by including those eigenoperators deep in the spectrum that are activated by the action of the perturbation on the initial eigenoperators.
This set of new operators could be determined using, e.g., Fermi's golden rule or the leakage amplitude of Eq.~\eqref{eq:resolvent}. 
Second, beyond the model considered here, LSS could be applied to richer and more hardware-specific error correction models with circuit-level descriptions of syndrome extraction, several independent noise sources, and additional coherent errors.

\section{Relation to other available methods}
\label{sec:comparison}

Several approaches have been developed to simplify time-dependent open quantum dynamics by constructing effective generators, casting to enlarged state spaces, or reducing the cost of the direct integration of the equations of motion.

When the timescale of the time-dependent perturbation is well separated from that characterizing the dominant processes of the system, a standard toolbox of approximations is available.
For example, a suitable change of frame can sometimes confine the explicit time dependence to a few counter-rotating terms.
If these terms oscillate rapidly compared to the relevant coupling strengths, the rotating-wave approximation justifies discarding them, yielding an accurate static effective generator~\cite{walls_quantum_2008}.
When stochastic noise decorrelates much faster than the intrinsic system dynamics, the Markov approximation captures its effects through effective time-independent jump operators~\cite{gardiner2004quantum}.
More systematically, recursive expressions such as the Magnus or cumulant expansions construct an effective static generator, of which the rotating-wave or the Markov approximations are the leading orders~\cite{magnus1954,Kubo1962,blanes2009,vanKampen1974a}.
However, resolving exponentially small rates may require very high orders to be accurate~\cite{carde2025}.

For structured quantum environments, hierarchical equations of motion~\cite{TanimuraKubo1989,tanimura2020heom}, pseudomode mappings, and chain mappings~\cite{tamascelli2018pseudomodes,somoza2019dampf,chin2010tedopa,tamascelli2019ttedopa} retain environmental memory through auxiliary operators or explicit environmental degrees of freedom.
For classical Gaussian noise with Lorentzian (damped-oscillator) correlations, as used above, sampling-free hierarchies of the Kubo--Anderson stochastic-Liouville type~\cite{Anderson1954,Kubo1954,kubo1963,Fox1978} play the same role. These methods trade the cost of averaging over noise realizations for a single deterministic equation in an augmented space, so their cost grows with the number and dimension of the coupled degrees of freedom and with the hierarchy depth. This scaling becomes particularly severe for driven, highly occupied bosonic modes~\cite{Weimer2021}. 
LSS instead keeps the per-realization propagators and places no restriction on the shape of $S(f)$.

Closer to the present setting are master equations for driven systems under classical non-Markovian noise~\cite{Groszkowski2023} and the filter-function formalism~\cite{cywinsky_may2008,Green2013,PazSilvaViola2014,Cerfontaine2021}, which express decoherence as an overlap between $S(f)$ and a control-dependent transfer function. The second-order Born approximation of App.~\ref{sec:bornapprox} is of this type. These approaches are perturbative in the noise strength, whereas LSS is not and treats deterministic drives and noise with the same machinery. 
\subsection{Projection-based model reduction}
LSS is a projection-based model-order reduction method~\cite{BennerGugercinWillcox2015}. The compression step is a proper orthogonal decomposition (POD)~\cite{Berkooz1993} where snapshots are assembled into a matrix and its dominant singular vectors are retained. 
Standard POD, however, takes snapshots of the state along solution trajectories, and therefore requires solving the full problem first
and ties the basis to one initial condition and time horizon. LSS instead takes snapshots of the slow eigenoperators of the instantaneous Liouvillian, at the cost of a few diagonalizations of the snapshots, so
that the resulting basis is a property of the distribution of the perturbation rather than of any one trajectory.

Closer in spirit are Krylov-subspace approaches to the slow modes of a static
Liouvillian~\cite{MingantiHuybrechts2022}, and the observation, exploited in
Refs.~\cite{Christiansen2025,melo_variational_2025}, that eigenoperators at one
parameter value are well approximated by linear combinations of eigenoperators
at neighboring values~\cite{Herbst2022,Brehmer2023,Duguet2024}. LSS extends this observation from static parameter sweeps to time-dependent problems by replacing interpolation between fixed points with a residual-controlled basis valid across the sampled range. 

Closely related to LSS is the reduction of Ref.~\cite{casanova2026quantummodelreductionbased}, a non-perturbative method to identify the slowly decaying subspace, where time-dependent drives are treated by updating the subspace in time. LSS instead constructs a single basis valid over the whole range of the perturbation, which can be reused across different noise realizations.

Algebraic approaches provide another route to reduced open quantum dynamics. They construct reduced models that reproduce selected observables exactly while preserving Lindblad form~\cite{Grigoletto2025exactmodelreduction,grigoletto2025controlled}. Approximate extensions combine algebraic structure with spectral separation and perturbative arguments~\cite{grigoletto2026approximate}. LSS selects its subspace according to the slow spectrum, without requiring closure under operator multiplication. This allows the retained subspace to be chosen directly for the dynamics of interest, but does not impose a Lindblad structure on the projected generator.

\subsection{Other compression techniques}

Quantum-trajectory methods replace density-matrix evolution with an ensemble of stochastic pure-state trajectories, each of which is substantially cheaper to propagate~\cite{dalibard1992,dum1992,molmer1993,carmichael1993,plenio1998,daley2014,wiseman2010}.
Time-dependent noise and perturbations are naturally accommodated in this framework, since each trajectory is propagated with its own realization~\cite{Vicentini2019}.
However, accurately resolving exponentially rare events requires exponentially many quantum trajectories.
Importance sampling and large-deviation techniques partially mitigate this by preferentially sampling rare trajectories~\cite{garrahan2010,carollo2018,schile2018}.
However, extending these techniques to time-dependent quantum trajectories is not straightforward, since biasing towards rare events requires prior knowledge of the statistics one is trying to resolve.

A natural comparison for the LSS method is with low-rank methods for open quantum dynamics~\cite{LeBrisRouchon2013,LeBris2015,CaoLu2018,GravinaSavona2024,
AppeloCheng2025,McCaul2021}, which approximate the density matrix by a number ($r$) of pure-state components $\rho\simeq\sum_{k=1}^{r}p_k\ket{\psi_k}\bra{\psi_k}$ with $r$ much smaller than the Hilbert space dimension.
LSS is not a low-rank method, and the distinction is decisive for rare events. 
Low-rank methods truncate on probabilities $p_k$ while LSS truncates on decay rates and oscillation frequencies, irrespective of weight. 
In the metastable regime, the slow eigenoperators are linear combinations of density operators describing two macroscopically distinct branches~\cite{Minganti2018}. Rare switches between these branches, mediated by otherwise improbable intermediate states, give rise to an exponentially small average rate of population transfer (see, e.g., Refs.~\cite{ferrari2026chaos,ferrari2026erasure,Carde2026} for bistability in cat qubits).
Low-rank approximations risk missing these crucial states, unless dynamically adaptive and specifically optimized for this~\cite{pietro2026}.
LSS reduces the \emph{dimension} of the Liouville space in which the dynamics is propagated, while placing no constraint on the Hilbert-space rank of the operators within that subspace, thus correctly capturing rare events.

\section{Conclusion}

We have introduced Liouvillian spectral snapshots (LSS), a reduced-basis
method for propagating Lindblad dynamics subject to time-dependent (potentially stochastic) 
perturbations.
The reduced basis is built from the slow eigenoperators of the Liouvillian at
a few values of the perturbation, sampled from its distribution and selected
greedily, and is then reused for every noise realization. 
The accuracy of LSS is controlled by two distinct mechanisms. When the
perturbation is strong enough to rotate the slow subspace beyond the span of a
single snapshot, the residual test detects the deficiency and the greedy loop
appends the missing directions. The only error that the construction
does not control is genuine leakage into the discarded sector (cf.
Eq.~\eqref{eq:resolvent}), which is largest when the perturbation is resonant
with transitions out of the retained window. Nevertheless, such errors are
limited in dissipative quantum systems  as weight transferred
to the discarded modes decays at least at the rate $\gamma_{\rm rb}$ set by the
edge of the spectral window, limiting the accumulation of errors over long times. 

We have shown the usefulness of the LSS method both for a bosonic-mode description of quantum hardware and for a multi-qubit error correcting code. 
Indeed, we have benchmarked the method against numerically exact dynamics for a bistable Kerr resonator and for a repetition code subject to correlated noise. 
In both of these examples, the  LSS reduces the dimension of the space in which
the dynamics is propagated by at least an order of magnitude while attaining agreement with numerically exact solutions.

The results presented here also motivate several extensions of LSS.
First, the accuracy of the algorithm can be improved for applications that require high precision. 
For example, by a dynamical enrichment of the reduced basis by including the discarded eigenoperators that couple most strongly to the retained sector.
Second, while we have mainly focused on superconducting circuit and continuous error correction applications, LSS extends naturally to any open quantum system, provided it has the right separation of timescales.
In particular, the LSS method could be extended to capture the properties of systems described by discrete CPTP maps representing, e.g., quantum gates or measurements.

\section{Acknowledgements}

H.C. acknowledges support from the Novo Nordisk Foundation
grant NNF20OC0060019.

\appendix

\begin{onecolumn}

\section{LSS Error Analysis}
\label{app:errors}

In the following, we bound the trace distance between the exact density matrix and the density matrix found from the LSS reduced basis. Let $\rho_{\rm rb}(t)=\mathcal U^{\rm LSS}_{t,0}\rho(0)$ be the reduced
evolution of an initial state $\rho(0)$ that lies in the reduced subspace, $\mathcal P^{\rm rb}\rho(0)=\rho(0)$, and
$\rho_{\rm ex}(t)=\mathcal U_{t,0}\rho(0)$ the exact one. Consider the state $\mathcal U_{T,t}\mathcal U_{t,0}^{\rm LSS}\rho(0)$ found by evolving the initial state with the LSS propagator up to a time $t$ and then with the exact propagator from $t$ to a later time $T$. Differentiating this state with respect to $t$ we find

\begin{equation}
    \partial_t \qty[\mathcal U_{T,t}\,\mathcal U^{\rm LSS}_{t,0}\rho(0)] 
    =-\mathcal U_{T,t}\,
    \bigl[\LL(t)-\LL^{\rm rb}(t)\bigr]\,\rho_{\rm rb}(t).
\end{equation}
Integrating this expression with respect to $t$ from $0$ to $T$, and using $\mathcal L^{\rm rb}\rho_{\rm rb}=\mathcal P^{\rm rb}\mathcal L\rho_{\rm rb}$ gives the exact identity
\begin{equation}\label{eq:duhamel}
    \rho_{\rm ex}(T)-\rho_{\rm rb}(T)
    =\int_0^T\dd t\;\mathcal U_{T,t}\,
      (\mathds 1-\mathcal P^{\rm rb})\,\LL(t)\,\rho_{\rm rb}(t).
\end{equation}
The integrand represents the overlap of $\LL(t)\rho_{\rm rb}(t)$ with the discarded subspace $\mathds{1}-\mathcal P^{\rm rb}$. Since a CPTP map is a contraction in the trace norm~\cite{Rivas_2012},
$\norm{\mathcal U_{T,t}X}_1\leq\norm{X}_1$, a bound for the trace-distance, $\mathrm{TD}(t)=\frac{1}{2}||\rho_{\rm ex}(t)-\rho_{\rm rb}(t)||_1$ is given by 
\begin{equation}\label{eq:aposteriori}
    \mathrm{TD}(T)
    \;\leq\;\frac{1}{2}\int_0^T\dd t\;\norm{(\mathds 1-\mathcal P^{\rm rb})\LL(t)\rho_{\rm rb}(t)}_1.
\end{equation}
While this bound is exact, the weight that leaks out of the reduced basis is treated as persisting for the remainder of the evolution. It therefore does not exploit the dissipative nature of Liouvillian evolution, nor the potential quasi-adiabatic nature of the perturbation.

Expanding the LSS state $\rho_{\rm rb}(t)=\sum_n c_n(t)\tilde\rho_n(t)$ in terms of the normalized eigenoperators $\tilde\rho_n(t)$ of the instantaneous reduced Liouvillian, $\mathcal L^{\rm rb}(t)$, with eigenvalues $\tilde\lambda_n(t)$ and using that 
\begin{equation}
    (\mathds{1}-\mathcal P^{\rm rb})\mathcal L(t)\ket{\tilde\rho_n(t)} = [\mathcal L(t)-\tilde\lambda_n]\ket{\tilde\rho_n(t)},
\end{equation}
we find that 
\begin{equation}
    \mathrm{TD}(T) \leq \frac{\sqrt{N_\mathcal H}}{2}\int_0^T \dd t \sum_n |c_n(t)|r_n(t)  ,
\end{equation}
with $r_n=\bigl\lVert(\mathcal L(t)-\tilde\lambda_n)\ket{\tilde\rho_n}\bigr\rVert,$ defined as in Eq.~\eqref{eq:residual} with the Hilbert-Schmidt norm. The factor $\sqrt{N_{\mathcal H}}$ with $N_{\mathcal H}$ the closed Hilbert-space dimension, converts to the trace norm by $||X||_1\leq\sqrt{N_{\mathcal H}}||X||$. 
The greedy construction bounds the residuals $r_n<\epsilon_{\rm res}$ for all eigenpairs inside the validation window. If the perturbation is slow enough that it does not populate the state outside of the validation window, the bound reduces to
\begin{equation}
    \mathrm{TD}(T) \leq \frac{\sqrt{N_\mathcal H}\epsilon_{\rm res}}{2}\int_0^T \dd t \sum_n |c_n(t)|.
\end{equation}
For non-adiabatic noise, the leakage amplitude for leaving the reduced-basis window can be estimated within first-order perturbation theory in $\mathcal V$ by introducing the interaction-picture propagator $\widetilde{\mathcal U}_{t,t'}$ through
\begin{equation}
    \mathcal U_{t,t'} = \mathrm e^{\LL_0 t}\,\widetilde{\mathcal U}_{t,t'}\,
    \mathrm e^{-\LL_0 t'},
    \label{eq:lab_frame_propagator2}
\end{equation}
such that it satisfies 
$\partial_t\widetilde{\mathcal U}_{t,t'}=\widetilde{\mathcal V}(t)\widetilde{\mathcal U}_{t,t'}$
with $\widetilde{\mathcal V}(t)\equiv\mathrm e^{-\LL_0 t}\mathcal V(t)\mathrm e^{\LL_0 t}$, which is solved by $\widetilde{\mathcal U}_{t,t'}=\mathcal{T}\mathrm{e}^{\int_{t'}^{t}\dd s\,
    \widetilde{\mathcal V}(s)}$.
Expanding the time-ordered exponential to first order,
\begin{equation}
    \widetilde{\mathcal U}_{t,t'}=\mathds 1+\int_{t'}^{t}\dd s\,
    \widetilde{\mathcal V}(s)+\order{\mathcal V^2},
\end{equation}
and decomposing the perturbation into harmonics,
$\mathcal V(t)=\sum_m\mathcal V_m\mathrm e^{\mathrm i m\omega t}$, the matrix
element of the Schr\"{o}dinger-picture propagator between right, and left eigenoperators of $\LL_0$, $\rho_k$ and $\sigma_n$, reads
\begin{equation}\label{eq:A_leak}
    \mel{\sigma_n}{\mathcal U_{T,0}}{\rho_k}
    =\delta_{nk}\mathrm{e}^{\lambda_k T}+\sum_m\qty(\mathrm e^{(\lambda_k+\mathrm i m\omega)T}-\mathrm e^{\lambda_n T})
    \frac{\mel{\sigma_n}{\mathcal V_m}{\rho_k}}{\lambda_k+\mathrm i m\omega-\lambda_n}
    +\order{\mathcal V^2}.
\end{equation}
For $k$ retained and $n$ discarded, the
transfer amplitude is controlled by the resolvent
$|\mel{\sigma_n}{\mathcal V_m}{\rho_k}|/|\lambda_k+\mathrm i m\omega-\lambda_n|$,
which is Eq.~\eqref{eq:resolvent} of the main text. The transition amplitude is therefore small whenever no harmonic of the
perturbation is resonant with a retained-to-discarded transition,
i.e. the detuning $|m\omega-\mathrm{Im}(\lambda_n-\lambda_k)|$ is large compared with the coupling. Even on resonance the denominator is bounded from below
by $|\mathrm{Re}\,\lambda_k-\mathrm{Re}\,\lambda_n|\geq\gamma_{\rm rb}-|\mathrm{Re}\,\lambda_k|$,
so the transition amplitude into a discarded mode never exceeds the ratio of the coupling to
the distance between the retained mode and the edge of the window. For modes discarded by the frequency cutoff, $\omega_{\rm rb}$, no such bound holds and it should therefore always exceed the frequencies at which the perturbation has significant spectral weight. Even if these transitions occur, the dissipative nature of Liouvillian evolution means that the errors induced will decay exponentially. Specifically, if the perturbation induces leakage into the discarded sector, and is subsequently switched off for a time $\Delta t$, the intrinsic dissipation of the system means that this induced weight will vanish as $\mathrm{e}^{\Delta t\mathrm{Re}\,\lambda_{\rm rb}}$. This is in stark contrast to closed systems where $\mathrm{Re}\lambda_{\rm rb}=0$ and one instead must rely on the cancellation of fast-oscillating terms.

\section{Noise Averaging}
\label{app:noiseaveraging}

To simulate the influence of noise, we need to average the propagator 
\begin{equation}
    \mathcal U_{T,0}[\eta] = \mathcal T \mathrm{e}^{\int_0^T\dd t\qty(\mathcal L_0+\eta(t)\mathcal V)},
\end{equation}
with generator $\mathcal L(t) = \mathcal L_0 + \eta(t)\mathcal V$, over noise realizations, $\eta$. 
Here, $\mathcal V=-i[\hat A,\cdot ]$  where $\hat A$ is a Hermitian operator e.g. $\hat A=\hat a^\dag \hat a$ representing noise in the detuning. The time-dependent coefficient $\eta(t)$ is assumed to be a Gaussian stochastic process with two-point correlation function
\begin{equation}\label{eq:psddef_app}
\langle \eta(t)\,\eta(t') \rangle_S
    = \int_{f_{\rm ir}}^{f_{\rm uv}} \dd f\, S(f)\, \cos\qty(2\pi f[t-t']).
\end{equation}
as in Eq.~\eqref{eq:psddef}. We introduce the two-sided PSD $S_2(f) = \frac{1}{2}S(|f|)$ for $f_{\rm ir}<|f|<f_{\rm uv}$ and $S_2(f)=0$ otherwise, so that $\langle \eta(t)\eta(t')\rangle_S=\int_{-\infty}^{\infty}\dd f S_2(f)\mathrm{e}^{-2\pi if(t-t')}$. We can write the stochastic trajectories in their spectral decomposition as~\cite{TimmerKoenig1995} 
\begin{equation}
        \eta(t)  = \int_{-\infty}^\infty\dd f \mathrm{e}^{2\pi i f t } \sqrt{S_2(f)}\xi(f)
\end{equation}
where $\xi(f)$ is a complex stochastic process with correlation function $\expval{\xi^*(f)\xi(f')}=\delta(f-f')$ and $\xi^*(f)=\xi(-f)$ which ensures that $\eta(t)$ is real. Starting from an initial state $\rho_0$ at $t=0$, the noise averaged density matrix at time $T$ can be written as 
\begin{equation}\label{eq:rhotimeevolve}
    \na\rho(T) = \langle \mathcal U_{T,0}[\eta]\rho_0\rangle_S= \na{\mathcal U}_{T,0}\rho_0 ,
\end{equation}
with $\na{\mathcal U}_{T,0}=\expval{\mathcal U_{T,0}[\eta]}_S$. The noise-averaged eigenoperators and rates are then found from the spectral decomposition of $\na{\mathcal U}_{T,0}$ as  
\begin{equation}
    \bar{\mathcal U}_{T,0} = \sum_n \vert \bar\rho_n \rangle \langle \bar\sigma_n \vert \mathrm{e}^{\bar\lambda_n T}
\end{equation}
where the eigenvalues are again ordered according to their real parts such that $0=\mathrm{Re}\na\lambda_0\geq \mathrm{Re}\na\lambda_1\geq\ldots$.\footnote{The averaged quantum channel is guaranteed to be completely positive trace preserving (CPTP) since it is a convex combination of CPTP maps.}. From the averaged propagator we can extract the slowest relaxation rate as $-\mathrm{Re}\,\bar\lambda_1$. We can also extract the noise-averaged long-time expectation value of an observable defined as 
\begin{equation}
    \bar{\expval{O}} = \lim_{t\to\infty} \langle \hat O \vert \na{\mathcal U}_{t,0}\vert \rho_0 \rangle=\langle \hat O \vert \na{\rho}_{\rm ss}\rangle ,
\end{equation}
using that $\lim_{t\to\infty}\na{\mathcal U}_{t,0}=\vert \bar{\rho}_{\mathrm{ss}}\rangle \langle \mathds{1}\vert $ where $\na{\rho}_{\rm ss}$ is the steady-state eigenvector of $\na{\mathcal U}_{t,0}$ with unity eigenvalue.

\subsection{Born Approximation}
\label{sec:bornapprox}
Here we derive an expression for the noise-averaged propagator of the Liouvillian $\mathcal L(t) = \mathcal L_0+\eta(t)\mathcal V$, that is second order in the noise, $\eta(t)$. The propagator in the interaction picture, $\widetilde{\mathcal{U}}_{t,t'}$, is defined as
\begin{equation}
    \mathcal{U}_{t,t'} = e^{\LL_0 t}\,\widetilde{\mathcal{U}}_{t,t'}\,e^{-\LL_0 t'},
    \label{eq:lab_frame_propagator}
\end{equation}
and satisfies
$\partial_t\widetilde{\mathcal{U}}_{t,t'} = \eta(t)\widetilde{\mathcal{V}}(t)\widetilde{\mathcal{U}}_{t,t'}$ with $\widetilde{\mathcal{V}}(t) \equiv e^{-\LL_0 t}\mathcal{V}e^{\LL_0 t}$. It can therefore be written as the time-ordered exponential
\begin{align}
    \widetilde{\mathcal{U}}_{t,t'}[\delta]
    = \mathcal{T}\exp\!\left(\int_{t'}^t \dd t_1\,\eta(t_1)\,\widetilde{\mathcal{V}}(t_1)\right) =\mathds{1}+\sum_{n=1}^{\infty}\frac{1}{n!}
    \mathcal T 
    \prod_{k=1}^{n}\int_{t'}^{t}\dd t_k\left[\eta(t_k)\,\widetilde{\mathcal{V}}(t_k)\right].
\end{align}
 Since $\eta(t)$ is Gaussian, we use the cumulant expansion to write the noise averaged propagator in the interaction picture, $\bar{\widetilde{\mathcal{U}}}_{t,t'} = \expval{\widetilde{\mathcal U}_{t,t'}[\delta]}_S$ as 
\begin{align}\bar{\widetilde{\mathcal{U}}}_{t,t'}
    &= \mathcal{T}\exp\!\left(
    \frac{1}{2}\int_{t'}^t\dd t_1\int_{t'}^t\dd t_2\;
    C(t_1 - t_2)\,\widetilde{\mathcal{V}}(t_1)\widetilde{\mathcal{V}}(t_2)
    \right)
    \label{eq:noise_avg_propagator}
\end{align}
where $C(t_1-t_2)=\langle \eta(t_1)\eta(t_2)\rangle_S$ and $\mathcal T$ orders all $\tilde{\mathcal V}(t_i)$ across the expanded exponential. If the noise is white, that is $C(t_1-t_2)=\kappa_\phi \delta(t_1-t_2)$, the averaged propagator becomes $\bar{\widetilde{\mathcal{U}}}_{t,t'} = \mathcal T\exp{\frac{\kappa_\phi}{2}\int_{t'}^t\dd t_1 \widetilde{\mathcal V}^2(t_1)}$. If $\mathcal V=-i[\hat A,\cdot]$, with $\hat A$ a Hermitian operator, we can use $\frac{1}{2}\mathcal V^2=\mathcal D[\hat A]$ to write the propagator in the Schr\"odinger-picture simply as 
\begin{align}
    \bar {\mathcal U}_{t,t'}=\mathrm{e}^{(t-t')\mathcal L_{\rm eff}}, && \mathcal L_{\rm eff}=\mathcal L_0  + \kappa_\phi \mathcal D[\hat A]
\end{align}
from which the standard dephasing jump operator is recovered. 
If the noise has a finite correlation time and $[\mathcal L_0,\mathcal V]\neq 0$ no exact re-summation is possible in general. However, one can expand the propagator in Eq.~\eqref{eq:noise_avg_propagator} to second order in $\eta$ as $\na{\widetilde{\mathcal{U}}}_{T,0}=
    \mathds{1} + \na{\widetilde{\mathcal{U}}}^{(2)}_{T,0}+\mathcal{O}(C^2)$, where the second order correction, $\na{\widetilde{\mathcal{U}}}^{(2)}_{T,0}$, is  
\begin{align}\label{eq:g2_contourintegral}&\na{\widetilde{{\mathcal{U}}}}^{(2)}_{T,0}= \sum_{nkm} \ket{\rho_n}\int \dd f S_2(f)h_{nkm}(2\pi f) \mathcal V_{nk}\mathcal V_{km}\bra{\sigma_m}.
\end{align}
Here, $\mathcal V_{nk}=\langle \sigma_n|\mathcal V|\rho_k\rangle$ are the matrix elements of $\mathcal V$ with $\sigma_n$ and $\rho_n$, the left and right eigenvectors of the static Liouvillian, $\mathcal L_0$, and $h_{nkm}(\omega)$ is defined as \begin{align}\label{eq:effective_filter_function}
    h_{nkm}(\omega) &= \frac{\mathrm{e}^{T(\lambda_k-\lambda_n-i\omega)}-1}{(\lambda_k-i\omega-\lambda_n)(\lambda_k-i\omega-\lambda_m)}
    -\frac{\mathrm{e}^{T(\lambda_m-\lambda_n)}-1}{(\lambda_k-i\omega-\lambda_m)(\lambda_m-\lambda_n)},
\end{align}
where $\lambda_k$ are the eigenvalues of the static Liouvillian. From this, the lab-frame propagator is 
\begin{equation}
   \na {\mathcal U}_{T,0} = \mathrm{e}^{\mathcal L_0 T}\qty (\mathds 1 + \na{\widetilde{{\mathcal{U}}}}^{(2)}_{T,0})+\mathcal{O}(C^2). 
\end{equation}
from which all noise averaged quantities can be computed. 

\subsection{Numerical Computation of the Propagator}
\label{app:sm-strang}

Here we explain the details of the numerical scheme used to compute the time-dependent propagator presented in the main text and explain how the anti Hermiticity of the noise operator makes each noise propagation step an exact, norm-preserving update.  As in the main text, we consider a static Liouvillian perturbed by a classical noise process $\eta(t)$ that
couples to a Hermitian system operator such that $\mathcal V = -i[\hat A,\cdot]$ with $\hat A$ hermitian. This makes $\mathcal V$ anti-Hermitian, as can be seen from 
\begin{align}
  \langle \mathcal{V}[X]\vert Y\rangle
  &= \Tr\qty[ \{-i[\hat{A},X]\}^{\dagger} Y]
   = i\Tr\qty[(X^{\dagger}\hat{A} - \hat{A}X^{\dagger}) Y ]
   \nonumber\\[2pt]
  &= i\Tr\qty[X^{\dagger}(\hat{A}Y - Y\hat{A})]
   =  -\langle X\vert \mathcal{V}[Y]\rangle .
  \label{eq:sm-antiherm}
\end{align}
This means that $\mathcal V$ is normal with purely imaginary eigenvalues and diagonalized by a unitary. We now project to the reduced basis 
\begin{equation}
  \mathcal L_{0,\mathrm{rb}} = Q^{\dagger}\LL_0 Q,
  \qquad
  \mathcal V_{\mathrm{rb}} = Q^{\dagger}\mathcal{V}Q,
\end{equation}
where $Q$ is a rectangular matrix with orthonormal columns that span the snapshot eigenbasis. Since $Q$ is an isometry, $\mathcal V_{\rm rb}$ stays anti-Hermitian, so it can be diagonalized by the unitary $W$ such that $\mathcal V_{\rm rb} = WD W^\dag$ with $D=\mathrm{diag}(i\omega_1,i\omega_2,...)$. The full propagator can now be written in Trotterized form 
\begin{equation}
    \mathcal U_{T,0}^{\rm rb} = \prod_{j=0}^{M-1} \mathrm{e}^{\mathcal L_{\rm rb}(t_j)\Delta t},
\end{equation}
where in each factor, the generator is piecewise constant, $\mathcal L_{\rm rb}(t_j) = \mathcal L_{0,\rm rb} +\eta(t_j)\mathcal V_{\rm rb}$ and the time-step $\Delta t=T/M$ is set so it resolves the noise trajectories down to the ultra-violet cutoff $\Delta t\leq1/f_{\rm uv}$ and the reduced generator, $\Delta t||\mathcal L_{0,\mathrm{rb}}||\ll 1$. We split each step according to~\cite{strang1968}
\begin{equation}
    \mathrm{e}^{\mathcal L_{\rm rb}(t_j)\Delta t+\mathcal O(\Delta t^3)}\;=\; e^{\mathcal L_{0,\mathrm{rb}}\Delta t/2}\;
            W\,e^{\eta(t_j) \Delta t D}\,W^{\dagger}\;
            e^{\mathcal L_{0,\mathrm{rb}}\Delta t/2},
  \label{eq:sm-strang}
\end{equation}
where we neglect the $\mathcal O(\Delta t^3)$ term on the left-hand side which involves double commutators of $\mathcal V$ and $\mathcal L_0$. The power of this method is that in the bulk of the time domain, the computation of the propagator involves only the multiplication of $W^\dag \mathrm{e}^{\mathcal L_{0,\mathrm{rb}}\Delta t}W$, which is time-independent so needs to be computed only once, with $\mathrm{e}^{\eta(t_j)\Delta t D}$ which is a diagonal matrix.

\noindent\begin{minipage}{\linewidth}
\begin{algorithm}[H]
    \setstretch{1.15}

    \caption{Liouvillian spectral snapshots: greedy construction of $\mathcal P^{\rm rb}$}
    \label{alg:lss}

    \begin{flushleft}
    \noindent \textbf{Input:} time-dependent generator $\mathcal{L} (t)=\mathcal L_0+\mathcal V(t)$; distribution $p(\mathcal V)$ of the instantaneous perturbation;
     window $(\mathrm{Re}\,\lambda_{\rm rb}=-\gamma_{\rm rb},\mathrm{Im}\,\lambda_{\rm rb}=\omega_{\rm rb})$ (see Fig.~\ref{fig:scheme});
     tolerances $\epsilon_{\rm res}$, $\epsilon_{\rm SVD}$; rescaling factor $c$; number of consecutive passes $m_{\rm stop}$.\\
    \noindent\textbf{Output:}  the projector $\mathcal P^{\rm rb}$.
    \end{flushleft}
    \begin{algorithmic}[1]
    \State diagonalize $\mathcal L_{0}$
        inside the window set by $\lambda_{\rm rb}$
    \State $\mathcal B\gets {\rm orth} \{\ket{\mathds 1},\ket*{\rho^{(1)}_j}\}$,\quad
           $\mathcal P^{\rm rb}\gets\mathcal B\mathcal B^{\dagger}$,\quad $m\gets0$
    \For{$k=2,3,\dots$}
      \State draw $\mathcal V_k\sim p(\mathcal V)$,\quad
             $\mathcal L_k\gets\mathcal L_0+\mathcal V_k$
      \State diagonalize $\mathcal L^{\rm rb}_k=\mathcal P^{\rm rb}\mathcal L_k\mathcal P^{\rm rb}$
             $\rightarrow(\tilde\lambda_j,\ket{\tilde\rho_j})$
             \Comment{cheap operation: small matrix}
      \State $r_j\gets\lVert(\mathcal L_k-\tilde\lambda_j)\ket{\tilde\rho_j}\rVert$
             \Comment{cheap operation: one sparse product each}
      \If{$\max_j r_j<\epsilon_{\rm res}$} \Comment{For dense spectra: Run $j$ only over modes inside the validation window}
        \State $m\gets m+1$
        \If{$m=m_{\rm stop}$} \State \textbf{return} $\mathcal P^{\rm rb}$ \EndIf
      \Else
        \State diagonalize $\mathcal L_k$ inside the window;
               $\mathcal B\gets\mathcal B\cup\{\ket{\rho^{(k)}_j}\}$ \Comment{expensive operation}
    \State \label{ln:svd} SVD of $\mathcal B$;
           retain $\sigma_i/\sigma_1>\epsilon_{\rm SVD}$;
           rebuild $\mathcal P^{\rm rb}$
    \If{$\dim\mathcal P^{\rm rb}$ unchanged
        \textbf{and} $\epsilon_{\rm SVD}>\varepsilon_{\rm mach}$}
      \State $\epsilon_{\rm SVD}\gets\epsilon_{\rm SVD}/c$
      \Goto{ln:svd}
        \EndIf
        \State $m\gets0$
      \EndIf
    \EndFor
    \end{algorithmic}

\end{algorithm}
\end{minipage}

\vspace{-10pt}

\noindent\begin{minipage}{\linewidth}
\begin{algorithm}[H]
\caption{Liouvillian spectral snapshot: construction of the propagator}
\label{alg:propagator}
\setstretch{1.15}
\begin{flushleft}
\noindent \textbf{Input:} decomposition
         $\mathcal V(t)=\sum_{\mu}f_{\mu}(t)\,\mathcal V_{\mu}$ with fixed
         superoperators $\mathcal V_{\mu}$; window $\lambda_{\rm rb}$;
         number of noise realizations $R$ ($R=1$ for a deterministic perturbation); integration time $T$; list of times $t=0,\Delta t,2\Delta t,T-\Delta t$.\\
\noindent\textbf{Output:}  reduced propagator(s).
\end{flushleft}
\vspace{-5pt}
\begin{algorithmic}[1]
\Statex \emph{Offline stage --- performed once}
\State build $\mathcal P^{\rm rb}$ by greedy sampling of $p(\mathcal V)$
       \Comment{Call Algorithm~1}
\State $\mathcal L^{\rm rb}_0\gets\mathcal P^{\rm rb}\mathcal L_0\mathcal P^{\rm rb}$
\Comment{projected once, reused for every realization}
\State $\{\mathcal V^{\rm rb}_{\mu}\}\gets
          \{\mathcal P^{\rm rb}\mathcal V_{\mu}\mathcal P^{\rm rb} \,\, \forall\,\, \mu\}$
         \Comment{projected once, reused for every realization}
\vspace{5pt}
\Statex \emph{Online stage --- per realization}
\State $\bar{\mathcal U}^{\rm rb} \gets 0$
\For{$r=1,\dots,R$}
  \State generate the scalar coefficients $f_{\mu}(t)$ for this realization \Comment{See App. \ref{app:noiseaveraging}}
  \State $\mathcal U^{\rm rb}_{(r)}\gets\mathcal I$
  \For{$t=0,\Delta t,2\Delta t,\dots,T-\Delta t$}
    \State $\mathcal L^{\rm rb}(t)\gets\mathcal L^{\rm rb}_0
            +\sum_{\mu}f_{\mu}(t)\,\mathcal V^{\rm rb}_{\mu}$
           \Comment{scalar combination, no projection}
    \State $\mathcal U^{\rm rb}_{(r)}\gets
            \mathrm e^{\mathcal L^{\rm rb}(t)\Delta t}\,\mathcal U^{\rm rb}_{(r)}$
            \Comment{Speedup if $\mathcal V(t)=f(t)\mathcal V$ (see App.~\ref{app:sm-strang})}
  \EndFor
  \State 
  accumulate
         $\bar{\mathcal U}^{\rm rb}\gets\bar{\mathcal U}^{\rm rb}
          +\mathcal U_{(r)}^{\rm rb}/R$
\EndFor

\Statex

\end{algorithmic}
\end{algorithm}
\end{minipage}

\end{onecolumn}

\bibliographystyle{quantum}
\bibliography{refs}

\end{document}